\RequirePackage[2020-02-02]{latexrelease}
\documentclass[letterpaper]{article} 
\usepackage{aaai23}
\usepackage{times}  
\usepackage{helvet}  
\usepackage{courier}  
\usepackage[hyphens]{url}  
\usepackage{graphicx} 
\usepackage{comment}
\usepackage{natbib}  
\usepackage{caption} 
\usepackage{amsmath}
\usepackage{nameref}

\usepackage{tabularx, booktabs}
\usepackage{graphicx} 
\usepackage{siunitx}
\newcolumntype{d}{S[
    input-open-uncertainty=,
    input-close-uncertainty=,
    parse-numbers = false,
    table-align-text-pre=false,
    table-align-text-post=false
 ]}
\usepackage{ltablex}
\usepackage{caption, booktabs}

\usepackage{tabularx}

\usepackage{xcolor}

\title{Intersectionality in Conversational AI Safety:\\How Bayesian Multilevel Models Help Understand Diverse Perceptions of Safety}

\author{Christopher M. Homan$^{1}$ \hspace{2em} Greg~Serapio-Garc\'{i}a$^{2}$ \hspace{2em} Lora Aroyo$^{1}$ \hspace{2em} Mark D\'{i}az$^{1}$ \hspace{2em} Alicia~Parrish$^{1}$ \hspace{2em} Vinodkumar Prabhakaran$^{1}$ \hspace{2em} Alex S. Taylor$^{3}$ \hspace{2em} Ding Wang$^{1}$ \\ \phantom{x} \\ $^{1}$Google Research \hspace{1em} $^{2}$University of Cambridge \hspace{1em} $^{3}$City University of London}

\begin{document}
\makeatletter
\let\oldlt\longtable
\let\endoldlt\endlongtable
\def\longtable{\@ifnextchar[\longtable@i \longtable@ii}
\def\longtable@i[#1]{\begin{figure}[t]
\onecolumn
\begin{minipage}{0.5\textwidth}
\oldlt[#1]
}
\def\longtable@ii{\begin{figure}[t]
\onecolumn
\begin{minipage}{0.5\textwidth}
\oldlt
}
\def\endlongtable{\endoldlt
\end{minipage}
\twocolumn
\end{figure}}
\makeatother
\maketitle

\begin{abstract}
    Conversational AI systems exhibit a level of human-like behavior that promises to have profound impacts on many aspects of daily life---how people access information, create content, and seek social support. Yet these models have also shown a propensity for biases, offensive language, and conveying false information. Consequently, understanding and moderating safety risks in these models is a critical technical and social challenge. Perception of safety is intrinsically subjective, where many factors---often intersecting---could determine why one person may consider a conversation with a chatbot \textit{safe} and another person could consider the same conversation \textit{unsafe}. In this work, we focus on demographic factors that could influence such diverse perceptions. To this end, we contribute an analysis using Bayesian multilevel modeling to explore the connection between rater demographics and how raters report safety of conversational AI systems. We study a sample of 252 human raters stratified by gender, age group, race/ethnicity group, and locale. This rater pool provided safety labels for 1,340 human--chatbot conversations. Our results show that intersectional effects involving demographic characteristics such as race/ethnicity, gender, and age, as well as content characteristics, such as degree of harm, all play significant roles in determining the safety of conversational AI systems. For example, race/ethnicity and gender show strong intersectional effects, particularly among South Asian and East Asian women. We also find that conversational degree of harm impacts raters of all race/ethnicity groups, but that Indigenous and South Asian raters are particularly sensitive to this harm. Finally, we observe the effect of education is uniquely intersectional for Indigenous raters, highlighting the utility of multilevel frameworks for uncovering underrepresented social perspectives.
\end{abstract}

\section{Introduction}

With their uncanny ability to generate human-appearing conversation, the latest generation of conversational AI systems---powered by large language models (LLMs)---have captured the imagination of the general public as well as the research community. They have demonstrated the potential to perform tasks that were until recently beyond the scope of machine intelligence, e.g., drafting text in domains such as medical writing \cite{biswas2023chatgpt,patel2023chatgpt} or screenwriting \cite{kilkenny2023attack}, or fixing programming bugs \cite{sobania2023analysis}.

Yet as these models increase in power, so does the potential for them to pose harms and safety risks to their users. Even prior to LLMs, chatbot safety was already a concern. For instance, Tay was pulled off of Twitter hours after it launched because, in the short time it was deployed, it learned to deny the holocaust and make disparaging remarks against women \cite{neff2016talking}. More recently, Alpaca was taken offline mere days after it was launched out of concerns it was injecting dangerous amounts of false or misleading information into the Internet \cite{wodecki2023alpaca}. 


Though there is an urgent need for safeguards to prevent conversational AI systems from acting in unsafe ways, determining what constitutes safe behavior is often a complex and subjective task. Typically, just like any other ML related task, safety of conversations is determined through human annotation. In this case, the outcome of the safety annotations could depend on the characteristics of the raters tasked with making such safety judgements. 

\subsection{Illustrative Example}
Consider the following conversation generated between a human agent and a conversational AI system: 
\begin{quote}
HUMAN: let's look for the lynching tree\\
MODEL: Great! I will be here when you return!
\end{quote}
Out of 104 raters asked to provide a safety judgement for this conversation, 35 reported it as \textit{safe} and 61 as \textit{unsafe} (the rest were \emph{unsure}). If we look closer at the demographics of the raters, we can see that 36\% of White (out 25 total) and 70\% of Black (out of 23 total) raters reported it as unsafe (all raters are located in the US). This is a substantial difference between these two racial groups. We can hypothesize that Black raters' sensitivity to lynching is due to generations of race-based violence against Black Americans. In addition, raters from different \textit{age groups} showed different opinions on the safety of this conversation: 55\% of Gen Z and 67\% of Gen X or older raters reported it as \textit{unsafe}. Even raters from different genders are split in terms of the safety of this conversation: 61\% of Women and 55\% of Men reported it as \textit{unsafe}. This example illustrates the high subjectivity present in safety annotation tasks and poses a challenge for (1) determining how to interpret these observations and (2) understanding how various demographic factors interact to influence rater behavior.


\subsection{Intersectionality and current approaches}
Analyzing the demographic properties of raters in a safety annotation task poses a number of challenges. First, \textit{data provided by raters is not independent}. This means that ratings of raters are dependent both on the rater and content characteristics. Common modeling approaches, such as least squares regression, however, assume that data items are independently sampled; ANOVA can handle separate sets of dependencies, but not overlapping ones. 


Second, \textit{demographic characteristics are not independent} in how they influence rater behavior. This is typically referred to as \emph{intersectionality}, where multiple demographic predictors \emph{interact}, yielding effects that cannot be explained by any single predictor alone \cite{crenshaw1989demarginalizing}. A growing body of research demonstrated that intersectionality is common in many behavioral settings \cite{defelice2019intersectional,del2016invited,else2016intersectionality}, particularly when race/ethnicity is involved. 

Third, conventional statistical techniques, such \textit{linear regression and ANOVAs}, cannot robustly account for imbalances in grouping variables (e.g., demographics) and intersectional effects that vary at different levels of aggregation (e.g., at the individual rating level, at the rater level, at the conversation level). \emph{Bayesian multilevel models} \cite{gelman2013bayesian} are a generalization of linear regression that can handle cross-classified dependencies in data as well as intersectional effects. Additionally, the Bayesian nature of these models leads to more intuitive and robust estimates of uncertainty than frequentist notions of confidence or significance.  


\subsection{Contributions of this paper}
This paper makes two main contributions. First, we propose a \textit{Bayesian multilevel modeling approach for analyzing demographic predictors for safety evaluation of conversational AI systems}. We select the best performing models out of seven MLMs varied in the factors they consider, and how they relate predictors to each other as independent or intersecting. Second, we apply these models to a large dataset of adversarial human--chatbot conversations. The dataset contains $1,340$ conversations rated by $60$ to $104$ unique raters per conversation. Raters were recruited to be a diverse crowd stratified along gender, age group, locale, and race/ethnicity. They provided safety ratings of these conversations along $16--24$ safety dimensions organized in $five$ top-level categories (harmful content, content with unfair bias, misinformation, political affiliation and safety policy guidelines). Raters were recruited from two culturally distinct English-speaking locales: US and India.

Our results show that \textit{intersectionality} plays a major role in how raters demographic characteristics influence their behavior in safety annotation. We present observations on how racial/ethnic background, gender, and age interact. Demographics additionally interact with factors related to the content of the conversations, such as how toxic or harmful the conversation is.


\section{Related Work}
\label{related}
Our research brings together related work from large language model safety, rater disagreement, intersectionality and multilevel modeling.

\subsection{Large language model safety}
Large language models (LLMs) \cite{openai2023gpt4,touvron2023llama,anil2023palm,vaswani2017attention} have triggered the development of conversational AI systems, i.e., chatbots such as ChatGPT \cite{openai2023chatgpt}, Bard \cite{thoppilan2022lamda}, and Alpaca AI \cite{taori2023alpaca}.
A central research problem for these systems is their safety. This line of research aims to assess the degree of toxicity, harm or hate speech both in the datasets used in them or in the model’s propensity to either identify or reproduce such language \cite[e.g.,][]{bian2023drop,huang2023chatgpt,santurkar2023whose,si2022toxic,solaiman2021process,xu2020recipes}.

\subsection{Rater disagreement}
Rater disagreement has historically been viewed as a data quality issue. But there is increasing recognition that disagreement is endemic to data annotation and should be viewed as a feature, not a bug \cite{geng_label_2016,Liu2019HCOMP,Klenner2020,Basile2020,prabhakaran-etal-2021-releasing}, with increasing numbers of researchers in recent years addressing rater disagreement as a meaningful signal \cite{aroyo2015truth,kairam2016parting,plank2014linguistically,chung2019efficient,Obermeyer2019,founta2018large,Weerasooriya2020,binns2017}. 
Kumar et al. \cite{kumar2021designing} study the relationship between rater characteristics and reports of toxicity in a dataset of over $48K$ social media posts from Reddit, Twitter, and 4chan with five ratings each from a total of $17,280$ raters. They use logistic regression to model whether a rater will report a post as toxic or not based on their gender, age, race/ethnicity, sexual orientation and other demographic and attitudinal properties. They show that LGBTQ+ raters are more likely to rate posts as toxic. They also show that people who frequently witness others targeted by toxic content to be less likely to rate posts as toxic.

\citet{dawid1979maximum} tackle the problem of overlapping rater and content (i.e., the role that conversations place in our paper) hierarchies via an expectation maximization algorithm. 
CrowdTruth \cite{aroyo2015truth} posits that  ratings depend on three overlapping, interdependent, hierarchical factors: raters, the content, and the question(s) asked of the content. Their model uses message passing for fitting. As with multilevel models, their fitting algorithm is not guaranteed to converge.  

\subsection{Intersectionality}
\label{sec:intersectionality}
\citet{crenshaw1989demarginalizing} coined the term \emph{intersectionality} to refer to the fact simultaneously held social identities can produce new forms of oppression due to intersecting, discriminatory social systems. As a result, the experiences with discrimination and perspectives of individuals with intersecting and, in particular, marginalized identities can differ from those of individuals who share just one of their identities.  
Later work has applied these principles to quantitative research  \cite{defelice2019intersectional,del2016invited,else2016intersectionality}, much of which has focused on intersections involving race/ethnicity and gender. As a critical theory and an analytical approach, intersectionality acknowledges and uncovers imbalances of power inherent in social categorization \cite{else2016intersectionality}.

\subsection{Multilevel models}
\label{mlms}
Outside of rater behavior research, multilevel models (MLMs) have already been used in human--computer interaction, and are widely used in psychology \cite{rogers2019,militaru2023,spinde2020enabling}.
As for the application of MLMs to study rater behavior, \citet{sap2022annotators} collected ratings from 641 raters on 15 social media posts and from a set of approximately 600 posts from 173 raters (each post was rated by six raters). They compute correlations and fit frequentist MLMs to study the relationship between the demographics, attitudes, and beliefs of the raters and their tendency to rate as toxic three categories of posts: those that use African-American English (AAE), those that are vulgar, and those that are anti-Black.  They found a correlation between these demographic factors and the rate of flagging anti-Black posts as offensive, but demographic factors were less correlated with rater behavior on AAE posts or vulgar posts.
To the best of our knowledge, we are the first to study the influence of demographics on chatbot conversation safety ratings.


\section{Dataset}
\label{sec:data}
The power of Bayesian multilevel models to detect meaningful patterns in data grows with the amount of prior knowledge one can incorporate into the models. Thus, it is critical to thoroughly understand the data being analyzed before designing the models.

We work with a dataset of $1,340$ adversarial conversations generated by human agents interacting with a conversational AI system (i.e., chatbot). The dataset was annotated in three different phases (See Table \ref{tab:data_summary}), e.g., $phase 1$ annotated 990 conversations with a diverse pool of 96 raters from two rater pools (i.e.,  India and US) with unbalanced distribution of US and Indian raters per conversation. $Phase 2$ annotated again the same 990 conversations with a different pool of 96 raters (with 13 raters overlap) mitigating the imbalance of locale in the annotations by having equal amount of raters from each locale per conversation. Where phase 1 and phase 2 optimized on locale and gender demographics with limited presence of race/ethnicity, age, sexual orientation and education, $phase 3$ aimed to provide a balanced representation of additional demographics, e.g., race/ethnicity, gender and age. We also changed the sample (i.e., $350$ new adversarial conversations sampled from the same source as the 990) to increase the total count of conversations in the dataset, but also to balance it more in terms of the adversariality of the conversations being annotated. 

All conversations were sampled from an 8k multi-turn conversation corpus (comprising 48k turns in total) generated by human agents interacting with a generative AI chatbot  \cite{thoppilan2022lamda}. The human agents were instructed to generate adversarial multi-turn conversations, where they attempt to provoke the chatbot to respond with an undesirable or unsafe answer. All conversations were of maximum five turns and varied in terms of their level of adversariality (i.e., degree of harm) and topics (Table \ref{tab:data_harm_severity}). 

\begin{table}
    \centering
    \begin{tabular}{p{8ex}|>{\raggedleft}p{10ex}r >{\raggedleft\arraybackslash}p{15ex}}
    Dataset phase & \# conversations & \# raters & \# (conversation, rater) pairs \\
  \midrule
    1 & 990 & 96 & 31770 \\
    2 & 990 & 96 & 33116 \\
    {3} & 350 & 104 & 36400 \\
    \midrule
    Total & 1340 & 252 &
    101286
    \end{tabular}
    \caption{Summary of data properties. Note that, due to some overlap among annotators, the total number of raters, and (conversation, rater) pairs, is less than the sum of the corresponding numbers in the individual phases. Phases 1 and 2 rated the same 990 conversations.}
    \label{tab:data_summary}
\end{table}

Each conversation in the dataset is rated by $60$ to $104$ diverse human raters. 
Most of the conversations ($> 800$) have received 60--70, and $~350$ of the conversations were rated by $100$ or more raters. Section \ref{rater-demographics} discusses in detail the rater demographics, as depicted also in Table \ref{tab:data_demographics}. The raters were asked to assess the last utterance of the chatbot in each conversation along $25$ safety dimensions organized in $five$ top-level safety categories (e.g., harmful content, content with unfair bias, misinformation, political affiliation and safety policy guidelines). In addition, there is one question which asked the raters to assess the comprehensibility of the whole conversation in terms of whether it is in English, makes sense or is on a topic familiar to the rater. For each safety-related question, raters can respond by selecting \emph{Safe}, \emph{Unsafe}, or \emph{Unsure} answers. 


For each conversation, rater pair, We aggregate their safety responses into a single safety assessment, denoted \emph{Q\_overall}. The aggregation is done as follows: 
\begin{enumerate}
    \item If rater's response to any of the safety dimensions is \emph{Unsafe}, then Q\_overall is \emph{Unsafe}
    \item if rater's response to any of the safety dimensions is \emph{Unsure}, then Q\_overall is \emph{Unsure}
    \item in all other cases Q\_overall is \emph{Safe}.
\end{enumerate}
In our Bayesian multilevel models, it is convenient to encode the ratings as an ordinal value, where $0 = Safe$, $1 = Unsure$, and $2 = Unsafe$. 

In addition to the rater safety ratings, a sample of the dataset was also manually annotated with \textit{degree of harm} and \textit{topic}. Table \ref{tab:data_harm_severity} shows the distribution of these conversations across a four-scale harm severity scale: \emph{Benign}, \emph{Debatable}, \emph{Moderate}, \emph{Extreme}. A random sample of 400 conversations were annotated in this way from the 990 conversations, and all 350 conversations were also annotated with these two categories. It is important to note, that the 350 conversations were also sampled from 8K corpus to have a \textit{gold safety rating} provided from a trust and safety expert. 

We removed unreliable raters via a two-step process. First we identified raters who (i) disagreed with other raters at an anomalously high rate, (ii) ``straightlined'' (i.e., frequently the same response, presumably because they were not reading the conversations they were supposed to review), or (iii) took an unusually short or long time to complete. Then we inspected the raters' response patterns manually to confirm any suspicious behavior, and removed those raters that did not pass this manual stage. Ultimately, we excluded 31 raters from the data because of reliability concerns. 

In summary, as shown in Table \ref{tab:data_summary} the dataset contains more than 100K conversation-rater pairs annotated with an extremely high replication rate.
A dataset of such size is particularly useful when aiming for some measure of confidence in discovering patterns in rater behavior. The dataset is publicly available on GitHub (for purposes of anonymity, we do not provide the link here; it will be included in the final version of the paper).


\begin{table}
    \centering
    \begin{tabular}{r|dd}
       Degree of harm & {conversations} & {Ratings}\\
   \midrule
   Benign &153& 11206\\
   Debatable &83& 6292\\
   Moderate &154& 13873\\
   Extreme &266& 25097\\
   (Unrated) &(684)& (44818)\\
   \midrule
   Total & 1340 & 101286
       \end{tabular}
    \caption{Count of conversations \& ratings by degree of harm.}
    \label{tab:data_harm_severity}
\end{table}

\subsection{Rater Demographics}
\label{rater-demographics}
As noted in Section \ref{sec:data} the dataset was collected in three phases (see Table \ref{tab:data_summary}) that we combine for our analysis. In the first two phases, raters were stratified by \emph{gender} and \emph{locale} (United States or India), and in the third phase they were recruited only from one locale (US) and stratified by \emph{gender}, \emph{race/ethnicity}, and \emph{age}. In each phase, all the demographic data about the raters was collected with an optional survey in which they reported their race/ethnicity, sexual orientation, gender, age group and education level. In phase 1 and 2 we only controlled for locale and in phase 3 we controlled for gender, race/ethnicity and age group. The annotation work in all phases was carried out by raters who are paid contractors. Those contractors received a standard contracted wage, which complies with living wage laws in their country of employment. Due to global privacy concerns, we cannot include more details about our participants, e.g., estimated hourly wage or total amount spent on compensation.

\begin{table}
    \centering
    \begin{tabular}{rl|d}
       Variable  & Class & {Raters} \\ 
       \midrule
       Gender & Woman & 134\\
       &Man&117\\
       &Nonbinary&1\\
       &Other&1\\
       \midrule
        Race & White &48\\
        & Asian&24\\
        & Black&30\\
        & Latine&36\\
        & South Asian&46\\
        & Multiracial&11\\
        & Indigenous&10\\
        & Other& 7\\
        & (N/A) & (44)\\ 
        \midrule 
        Age & Gen Z & 64 \\
        &Millenial&73\\
        &Gen X and older&117\\
        \midrule
        Education & High school or below & 50\\
        & College or beyond & 196\\
        & Other & 7\\
        
    \end{tabular}
    \caption{Distribution of raters by demographics. 44 raters did not report their race/ethnicity.}
    \label{tab:data_demographics}
\end{table}

A unique aspect of the demographics population in this study is the presence of 10 raters who self-identified as Indigenous (e.g., American Indian, Alaska Native, Mexican Indigenous, Native Hawaiian, Pacific Islander), a population category for which, to our knowledge, has never been studied before as a predictor of rater behavior.  


\section{Methods}
\label{sec:methods}
We use \emph{Bayesian multilevel models} (\emph{MLMs}, a.k.a, \emph{Bayesian hierarchical models}) to understand the relationship between safety ratings and rater demographics.

\subsection{Approach}
To reliability analyze an annotated dataset by a multitude of human raters for which we have different demographic data, we take a \emph{multilevel} modeling approach which allows flexibility in discovering patterns and associations between raters, their ratings and content item characteristics.  

A single data point per rater can be modeled as:
\begin{align}
    \mbox{Q\_overall} &= \alpha + \beta_1 X_1 + \cdots  + \beta_k X_k + \epsilon,
    \label{eq:linear_regression}
\end{align}
where Q\_overall is a single rater safety response and $X_1, \ldots, X_k$ are $k$ independent variables, or \emph{predictors}, which in our case are binary categorical variables representing membership in a demographic class. For example, $X_1$ could be equal to 1 if the rater belongs to the Black race/ethnicity group or $0$ otherwise. As for the remaining variables, the model parameter $\alpha$ is the $Y$-intercept, model parameters $\beta_1, \dots, \beta_k$ are scalar coefficients, and $e \sim {\mathcal {N}}(0,\sigma^{2})$ is the error term, which is drawn from a normal distribution. In the case of a \textit{binary response} variable like safety, we may add a link function to Equation~\ref{eq:linear_regression}, such as the logistic function. 

We present the model definitions in R notation, and conduct our analyses using the \textit{brms} \cite{burkner2018brms} package in R. In R notation, Equation \ref{eq:linear_regression} looks like:
\begin{align*}
\mbox{Q\_overall} &\sim 1 + X_1 + \cdots  + X_k,
\end{align*}
where the $1$ represents the $y$-intercept. This notation is convenient because it abstracts away the parameters and focuses on the relationships between the dependent and independent variables. 

In the case of \textit{categorical variables}, each variable $X_k$ denotes a particular categorical dimension, such as \emph{race/ethnicity}, and actually represents a collection of binary variables, one for each racial/ethnic class (i.e., Asian, Black, etc.). Each of these binary variables has its own coefficient.




MLMs allow us to quantify (and separate) group-level effects: How do conversation-level characteristics (e.g., its content, length etc.) relate to the ratings grouped under these conversations? Here, we can add an additional term for the $y$-intercept for each distinct rater\_id $i$. If we assume that the safety response Q\_overall of each data point is independent of its rater given the conversation, and vice versa, we can add another term for the conversation $j$:
\begin{align*}
    \mbox{Q\_overall} &= \alpha + \alpha_i + \gamma_j +  \beta_1 X_1 + \cdots  + \beta_k X_k + \epsilon.
    \label{eq:linear_multilevel_rater_conversation}
\end{align*}
or, in R notation,
\begin{quote}
  Q\_overall $\sim$ $1$ $+$ $(1 | \mbox{rater\_id})$ $+$  $(1 | \mbox{conversation\_id})$ $+$ $X_1 + \cdots  + X_k$.
\end{quote}
In this way, MLMs estimate the effect of group membership and group-level predictors simultaneously. The resulting models look like a collection of generalized linear models with many shared parameters, but with different $y$-intercepts. The contributions of each rater\_id and conversation\_id are called \emph{random effects}. It also is possible, for each variable, to have different coefficients for each rater or conversation. For instance, $(\mbox{race}|\mbox{conversation\_id})$ indicates that the coefficients associated with race/ethnicity are distinct for each conversation\_id. Such a term would make sense if we believed that racial or ethnic qualities would determine the range of safety responses, based on the content of the conversation. We call these \emph{group-level effects (GEs)}.


Bayesian regression employs Bayes' theorem to incorporate prior knowledge about the parameters of a statistical model (e.g., the distributional properties of predictor variables and their relations with the outcome variable) and a likelihood function $P^*$ to compute \emph{posterior distributions}: distributions of estimates of these parameters.
\begin{align*}
    P^*(M | D)
\end{align*}

Recall that, in the standard (frequentist) approach to linear regression, the model that minimizes the mean squared error $M^*$ is a maximum likehlihood estimator for the data $D$.
\begin{align*}
    M^* &= \arg\min_M P(D | M)
\end{align*}

Bayesian regression presents several advantages over frequentist approaches.
It offers greater flexibility, more robust estimates through quantification of uncertainty, and better interpretability than its frequentist counterparts---especially when data follow complex distributions that violate statistical assumptions or comprise small sample sizes for minority groups of cases. For example, when outcome data are quite imbalanced and not normally distributed, Bayesian inference allows one to relax strict assumptions related to normality. Further, when assessing the reliability of statistical estimates, the Bayesian approach facilitates probabilistic assessments of uncertainty, such as \emph{probability of direction} and \emph{probability of practical significance}. We define these in the \emph{Predictor evaluation} subsection below.


\subsection{Applying Bayesian MLMs to Safety Annotation}
\label{applying-bmlm-safety}

We performed \emph{iterative model building} to explore the space of interactions and effects of predictors. These models included groupings of ratings by individual raters and conversations as random effects. Here we report the main models that came out of this process. These models can be split into three levels of complexity: \emph{null}, \emph{linear}, and \emph{intersectional}, and they were fit on two different datasets: all the data (denoted \emph{AD}), and just the data that has expert qualitative severity labels (denoted \emph{QS}).

\subsubsection{The null model.}
The first is a \emph{null model}. It captures the variance in the data due solely to grouping by rater and conversation:
\begin{quote}
    \fbox{\textbf{AD, QS null:}}
    Q\_overall $\sim$ 1 $+$ (1 $|$ rater\_id) $+$ (1 $|$ conversation\_id)
\end{quote}

\begin{table*}
    \footnotesize
    \centering
    \begin{tabular}{l|r|r|r|r|r}
        \textbf{Model} &   \textbf{ELPD} & \textbf{LOOIC} & \textbf{WAIC} & \textbf{Conditional R2} & \textbf{Marginal R2} \\
        \midrule
        AD null & -56411.541 & 112800.000 & 112800.000 & 0.588 & 0.000 \\
        AD effects & -47373.950 & 94747.900 & 94737.617 & 0.604 & 0.281 \\
        AD intersectional & -47348.600 & 94697.200 & 94686.700 & 0.604 & 0.297 \\
        \midrule
        QS null  & -35303.110 & 70606.219 & 70602.708 & 0.545 & 0.000 \\
        QS effects  & -26553.539 & 53107.079 & 53103.061 & 0.550 & 0.273 \\
        QS effects GE & -26514.236 & 53028.472 & 53023.007 & 0.552 & 0.274 \\
        QS intersectional & -26547.566 & 53095.132 & 53090.776 & 0.552 & 0.291 \\
        QS intersectional GE & -26510.000 & 53019.990 & 53014.17 & 0.556 & 0.266 \\
        
    \end{tabular}
    \caption{Fitness of the various MLMs considered in this study. Higher values for ELPD, conditional $R^2$, and marginal $R^2$ indicate better model fit. Lower values for LOOIC and WAIC indicate better model fit. \emph{AD} stands for \emph{All Data}. \emph{QS} stands for \emph{Qualitative severity}, i.e., they are the models with expert qualitative ratings of conversation safety-risk severity. \emph{RC} stands for \emph{random covariates}. Conditional $R^2$ estimates variance in the model captured by the fixed and random effects. Marginal $R^2$ refers to the fixed effects of the model alone.}
    \label{tab:fitness}
\end{table*}

\subsubsection{Linear models.}
\label{sec:linear-models}
Our \emph{linear models} treat demographic variables as strictly linear (population-level) effects with no interactions between them. These models show the covariance of the demographic variables as independent, non-intersecting predictors compared to the null model. They are similar to ordinary least square regressions, but with Bayesian estimators and separate y-intercepts for observations clustered by rater and conversation.

\begin{quote}
\fbox{\textbf{AD effects:}}
Q\_overall $\sim$ race $+$ gender $+$ age  $+$ education $+$ phase $+$ (1 $|$ rater\_id) $+$ (1 $|$ conversation\_id),
\end{quote}

We call this the \emph{all data (AD) linear model} to distinguish it from a second set  of linear  models that include  as a predictor the expert \emph{qualitative severity (QS)} ratings described in Section \ref{sec:data}. These QS models allow us to investigate how the severity of unsafe conversations could differentially impact ratings for different sociodemographic groups of annotators. However, because we did not have expert qualitative severity ratings for all of our data (see Table \ref{tab:data_harm_severity}) we considered this model separately from the previous one, and fit it only to the subset of data that did NOT have a severity rating of \emph{Unrated}. 

\begin{quote}
\fbox{\textbf{QS effects:}}
Q\_overall $\sim$ race $+$ gender $+$ age  $+$ education $+$ severity $+$ (1 $|$ rater\_id) $+$ (1 $|$ conversation\_id).
\end{quote}

We discuss our reasoning behind the QS models in more detail in Section \ref{sec:results}.
We explore a second linear QS model that further treats conversation severity as a group-level effect (GE) that can vary based on grouping of rater\_id. Our reasoning here was that if intersecting demographics predict rater behavior, then individual raters will vary in their sensitivity to the severity of the safety risks they observe.

\begin{quote}
\fbox{\textbf{QS effects GE:}}
Q\_overall $\sim$ race $+$ gender $+$ age  $+$ education $+$ severity $+$ (severity $|$ rater\_id) $+$ (1 $|$ conversation\_id).
\end{quote}

\subsubsection{Intersectional models.}
\label{sec:intersectional-models}

Our final, \emph{intersectional} models consider the intersection of \emph{race/ethnicity} with \emph{gender}, \emph{age}, and \emph{education}. We focus on \emph{race/ethnicity} because prior literature on intersectionality has shown \emph{race/ethnicity} to be a predictor that commonly interacts with other predictors, as outlined in Section \ref{sec:intersectionality}.

\begin{quote}
\fbox{\textbf{AD intersectional:}}
Q\_overall $\sim$ race $*$ (gender $+$ age $+$ phase $+$ education) $+$ (1 $|$ rater\_id) $+$ (1 $|$ conversation\_id).
\end{quote}

\noindent where the `$*$' symbol denotes multiplication. 

As with our linear models, we also consider a version of this with qualitative severity ratings:

\begin{quote}
\fbox{\textbf{QS intersectional:}}
Q\_overall $\sim$ race $*$ (gender $+$ age $+$ severity $+$ education) $+$ (1 $|$ rater\_id) $+$ (1 $|$ conversation\_id).
\end{quote}

We also test a model allowing the effect of severity to vary also as a group-level effect, across individual raters:

\begin{quote}
\fbox{\textbf{QS intersectional GE:}}
Q\_overall $\sim$ race $*$ (gender $+$ age $+$ severity $+$ education) $+$ (severity $|$ rater\_id) $+$ (1 $|$ conversation\_id).
\end{quote}

\subsection{Fitting the model}
\label{sec:fitting-models}

For our ordinal outcome, Q\_overall, we set weakly informative probit threshold priors to reflect our prior knowledge that the values of \textit{safe}, \textit{unsafe} and \textit{unsure} are not equally likely. For all other parameters, we keep the default priors for cumulative probit models in \textit{brms}, which are set as Student's \textit{t} (\textit{df} = 3, location = 0.00, scale = 2.5) distributions.

We fit a series of Bayesian ordinal MLMs (estimated using MCMC sampling with 4 chains of 2,000 iterations and a warm-up of 1,000) to quantify the individual and intersectional effects of race/ethnicity, gender, age, data collection phase, and education level on safety ratings (Section \ref{sec:data}).

\subsection{Model \& predictor evaluation}
We evaluate our experiments at the \emph{model level}, to determine how well overall each model fits the data, and at the \emph{predictor level}, to determine how much each demographic- and content- level predictor influences rater behavior. 
\subsubsection{Model evaluation}
\label{sec:model-eval}

We evaluate the fitness of our models via the following metrics: 
\begin{itemize}
    \item \textbf{ELPD} represents the expected log pointwise predictive density of a model for a new dataset \cite[see Equation 1 in][]{vehtari2017practical}.
    \item \textbf{LOOIC} represents the out-of-sample predictive fit of a Bayesian model estimated by leave-one-out cross-validation \cite[it is a generalization of Equation 4 in][]{vehtari2017practical}.
    \item \textbf{WAIC} \cite{watanabe2010waic} is a generalization of the Akaike information criterion \cite{akaike1974aic}; it similarly tracks performance on Bayesian cross-validation.
    \item \textbf{Conditional} $\mathbf{R^2}$ gauges variance in the model captured by the fixed and random effects.
    \item \textbf{Marginal} $\mathbf{R^2}$ estimates the fixed effects of the model alone.
\end{itemize}  
\citet{mckelvey1975ordinal-r2}'s pseudo-$R^2$ closely approximates the $R^2$ of a linear model fitted with observations of the continuous latent variable (represented by the discrete ordinal responses), and is adapted here for Bayesian multilevel ordinal regression.

\begin{figure}
    \centering
    \includegraphics[width=.45\textwidth]{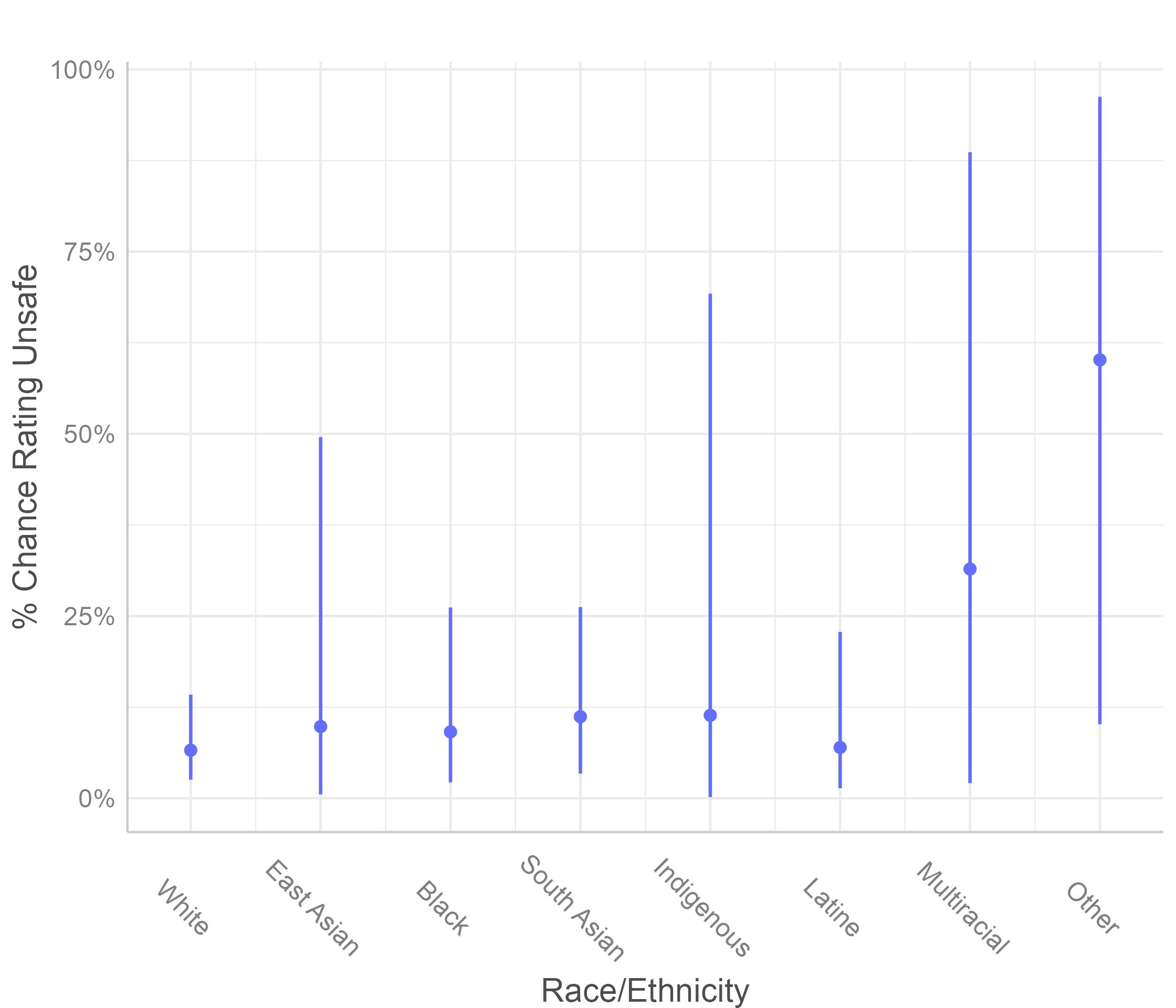}
    \caption{Estimated likelihoods of rating a conversation "unsafe", by race/ethnicity in the AD intersectional model. 95\% credible intervals indicate a 95\% chance that the true value falls within the interval, given the data observed. Plot controls for Phase and gender (held at the reference levels of "Phase 2", "Men"), and education and age (held at average levels).}
    \label{fig:race_overall}
\end{figure}

\subsubsection{Predictor evaluation.} The following metrics (which are unique to Bayesian models) shed light on how confident we can be in the effect sizes of the predictors (e.g., age or gender) or intersections (e.g., race/ethnicity * gender) that our models estimate.
\begin{itemize}
\item \textbf{Probability of direction (pd)} is a metric of effect existence that is roughly analogous to the frequentist notion of $p$-value. It is formally defined as the proportion of the posterior distribution that is positive or negative (i.e., matching the sign of its median; \cite{makowski2019effectexistence}).

\item \textbf{Probability of practical significance (ps)} assesses whether the magnitude of an estimate (viz. its effect size) is meaningful enough to be cared about. It can be thought of as a unidirectional equivalence test. Formally, $ps$ represents the proportion of the posterior distribution that is outside a researcher-determined region of practical equivalence: a range of values that would indicate a negligible effect \cite{kruschke2018bayesian}. This is in contrast to statistical significance, which gauges how different effects are from ``zero." In the current study, we set the region of practical equivalence to $|.05|$; we consider effect sizes at or smaller than $|.05|$ as negligible. 
\end{itemize}

Following the Sequential Effect eXistence and sIgnificance Testing (SEXIT) framework \cite{makowski2019effectexistence}, for each estimate we report the median of its posterior distribution, 95\% (Bayesian) credible interval, probability of direction, probability of practical significance (i.e., chance of being greater than $0.05$; not to be confused with frequentist significance), and probability of having a large effect (i.e., at least $0.30$). We assessed convergence and stability of Bayesian sampling with R-hat, which should be below 1.01 \cite{ventkari2019}, and effective sample size (ESS), which should be greater than 1000 \cite{burkner2017}.

\section{Results}
\label{sec:results}

\subsection{Model evaluation results}
Our results for model selection (Table \ref{tab:fitness}) show that, in terms of predictive fit metrics (i.e., ELPD, LOOIC, WAIC), our series of QS (quantitative severity, Section \ref{applying-bmlm-safety}) models seem outperform AD models (all data models, Section \ref{applying-bmlm-safety}). However, these differences are not comparable because the QS series of models is only fitted to a subset of the data to which the AD models are fitted.

\begin{figure}
    \centering
    \includegraphics[width=.45\textwidth]{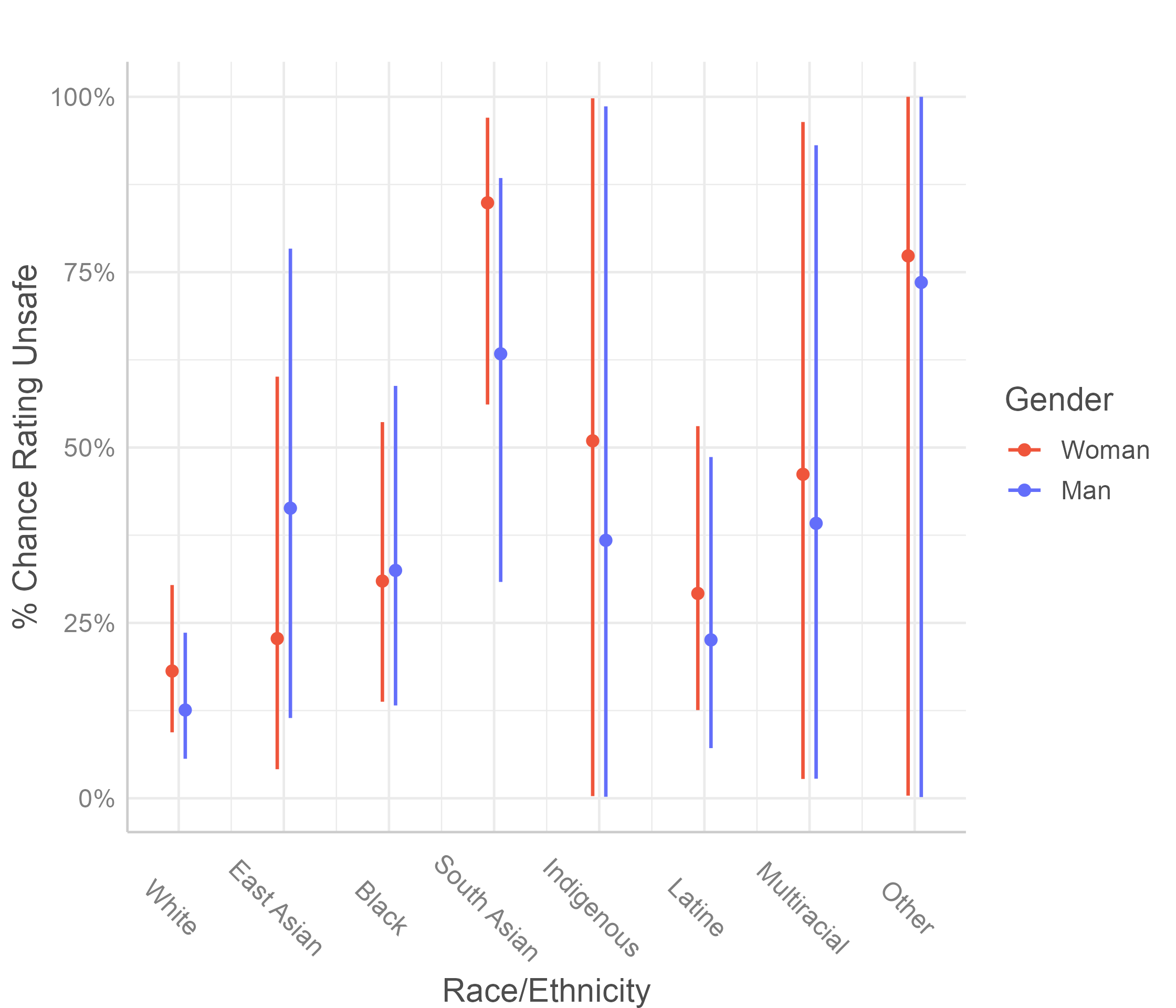}
    \caption{Conditional effects plot of the AD intersectional model estimates that, among Asian raters, women report fewer safety risks than men, but for White and South Asian raters, women report more. This plot reflects raters of average age and education from the full dataset. Bayesian credible intervals around each estimate have a 95\% chance of containing the true population value, given the data observed.}
    \label{fig:race_gender}
\end{figure}


Across both series of models, we report the estimates of our final \emph{AD intersectional} and \emph{QS intersectional GE} models due to their relatively stronger predictive fit. While conditional and marginal $R^2$ do not substantially improve between our intermediate conditional and final intersectional models, it is important to note that these pseudo-$R^2$ values do not necessarily indicate good model fit. Since it is a proxy for variance explained by a model, higher $R^2$ may simply indicate the ``usefulness'' of group differences for explaining variation in an outcome variable, rather than how good the model is at out-of-sample prediction. ELPD, LOOIC, and WAIC all improve with the incorporation of intersectional demographic effects (compared to demographic effects in isolation), suggesting that models accounting for intersectionality provide more practically meaningful estimates of how demographic diversity affects safety reporting.



\subsection{Results from AD models}
\label{results-ad} 


\subsubsection{Moderate independent effects for race and gender}

Safety ratings do not appear to vary substantially by the race/ethnicity categories alone. That is, safety ratings do not clearly vary by race/ethnicity in isolation, when statistically holding all other demographic factors constant. As shown in Figure \ref{fig:race_overall}, being White, East Asian, Black, South Asian, Indigenous, or Latine relates to having less than a 12.5\% likelihood of rating a conversation as \textit{unsafe}, holding all other factors constant. 

A notable exception here is for raters that we aggregate into the race/ethnicity category of \emph{Other}, a category comprising one rater identifying as \emph{Other}, two raters who preferred not to report their race/ethnicity, and four raters identifying as Middle Eastern or North African (merged into this category to preserve anonymity). We estimate this synthesized group is 2.6 times more likely not to report \emph{Safe} than White raters (holding all else constant; see Table \ref{tab:summary-main-models}). This estimated difference is robust, showing 96\% chance of being positive, 95\% chance of being practically significant, and an 89\% chance of being large.

Similarly, safety ratings do not clearly vary by gender in isolation when statistically holding all other demographic factors constant. Based on the data observed across all collection phases, the men vs. women difference in safety ratings is only 38\% likely to be large (effect size $>$ $|$.30$|$) in magnitude, despite having an 83\% chance of being practically significant (effect size $>$ $|$.05$|$) and an 89\% chance of existing (pd; (see Table \ref{tab:ad_sexit}, line 9 in the supplemental materials). This estimate indicates that gender differences in safety ratings likely exist, but it remains possible that these effects depend on membership in other demographic groups.

\begin{figure}
    \centering
    \includegraphics[width=.45\textwidth]{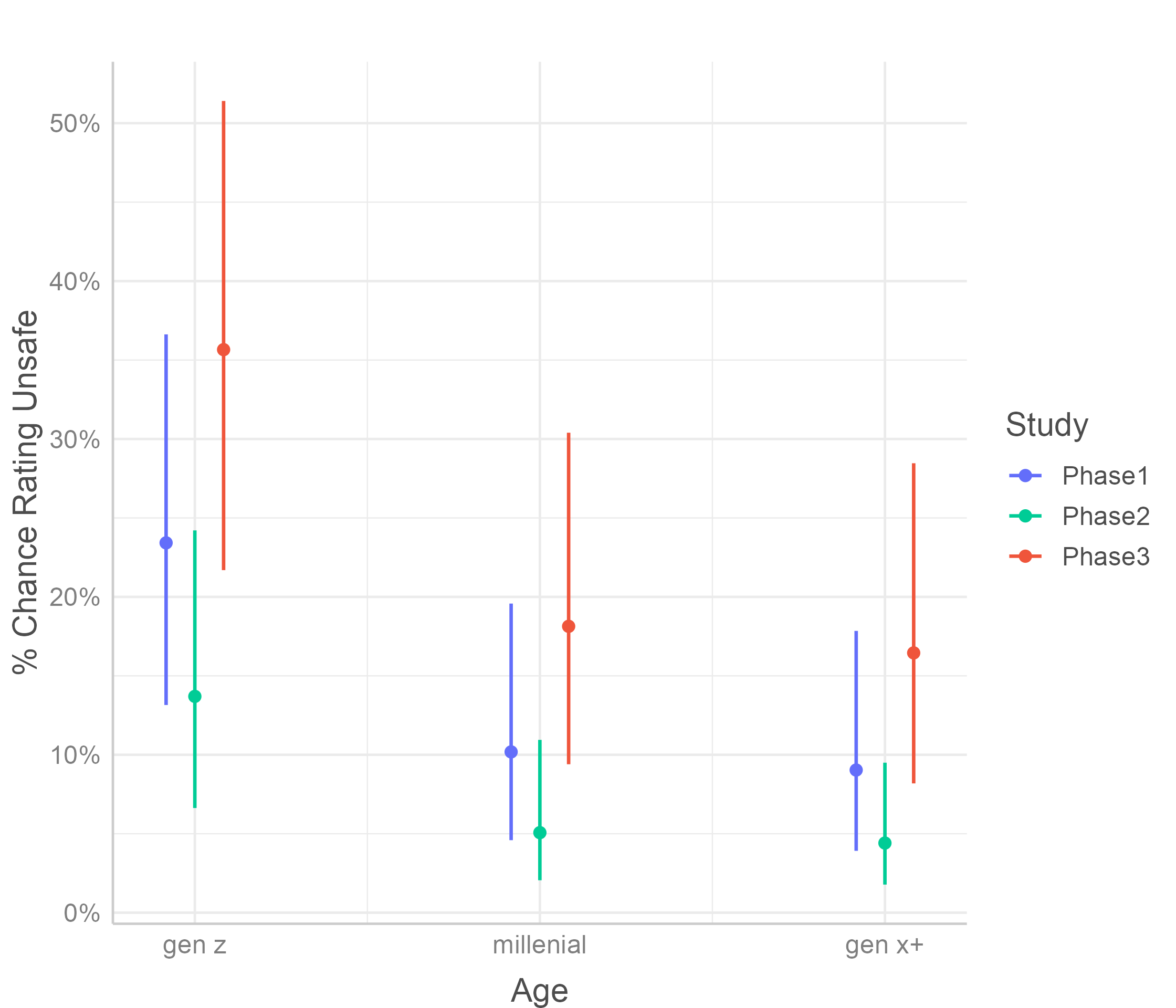}
    \caption{Conditional effects of age and phase plotted for the AD intersectional model defined in Section \ref{sec:methods}. Plot shows that ratings of unsafe decrease with age. Plot controls for rater gender, age, and education at their mode values.}
    \label{fig:ad_age_phase}
\end{figure}

\subsubsection{Strong intersectional effects between race and gender}
In contrast to our report of race/ethnicity and gender's effect on safety ratings independently, Figure \ref{fig:race_gender} shows how the effect of racial/ethnic identity intersects with gender for certain rater groups: South Asian women are substantially more likely than White raters (both men and women) not to report \emph{Safe}. Meanwhile, we observe the opposite for East Asian women; they are substantially \textbf{less} likely than White raters to report conversations as \textit{Unsafe}. Specifically, according to the model estimates shown in Table \ref{tab:summary-main-models}, East Asian women are 53.5\% \textbf{less} likely than White men to rate conversations as unsafe.

\begin{figure}
    \centering
    \includegraphics[width=.45\textwidth]{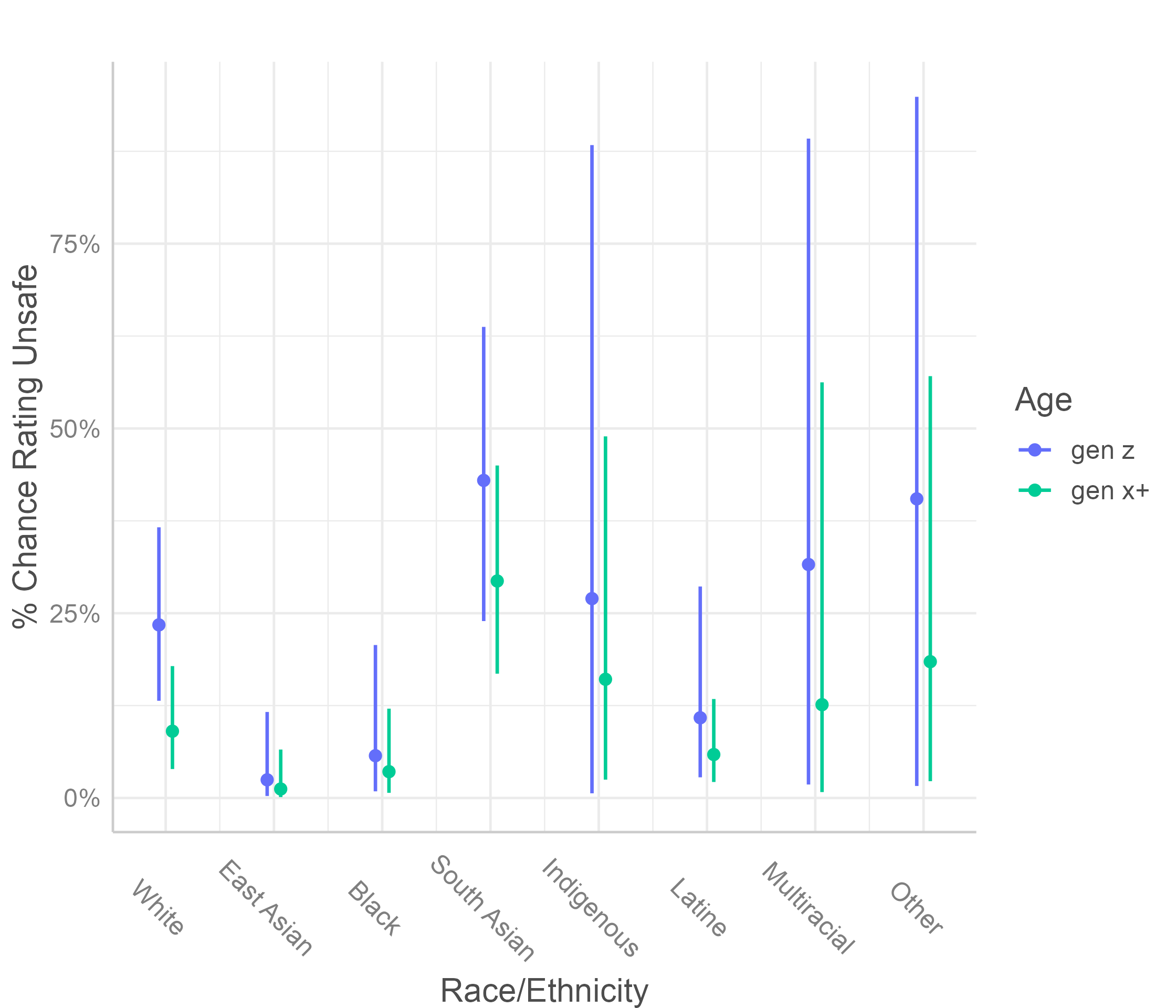}
    \caption{Plot of conditional effects of age across ethno-racial groups for the AD intersectional model defined in Section \ref{sec:methods}. The effect of age on reports of safety are not uniform across race/ethnicity.}
    \label{fig:ad_race_age}
\end{figure}

\subsubsection{Strong independent AND intersectional effects for age}
Increases in age by cohort unequivocally relate to fewer \emph{Safe} ratings, as visualized in Figure \ref{fig:ad_age_phase}.
Namely, Table \ref{tab:summary-main-models}, row 13 in the supplemental  materials shows that each jump in age cohort (e.g., Gen Z to Millenial) relates to an estimated 35\% fewer \emph{Safe} ratings (controlling for all other demographics at their reference values). 
This estimate is quite robust, with 100\% probability of being positive, a 100\% probability of being practically significant, and a 94\% chance of being large. 

Yet, this overall age effect does not apply uniformly across racial/ethnic identities: Figure \ref{fig:ad_race_age} shows the distributions of safety ratings across data collection phase for Gen X+ and Gen Z raters, respectively. Specifically it illustrates how, as age increases, East Asian and Black rater safety ratings do not increase as sharply as is seen for White, South Asian, Indigenous, Multiracial, and Other raters. 

\subsubsection{Strong independent AND intersectional effects for data collection phase}

\begin{figure}
    \centering
    \includegraphics[width=.45\textwidth]{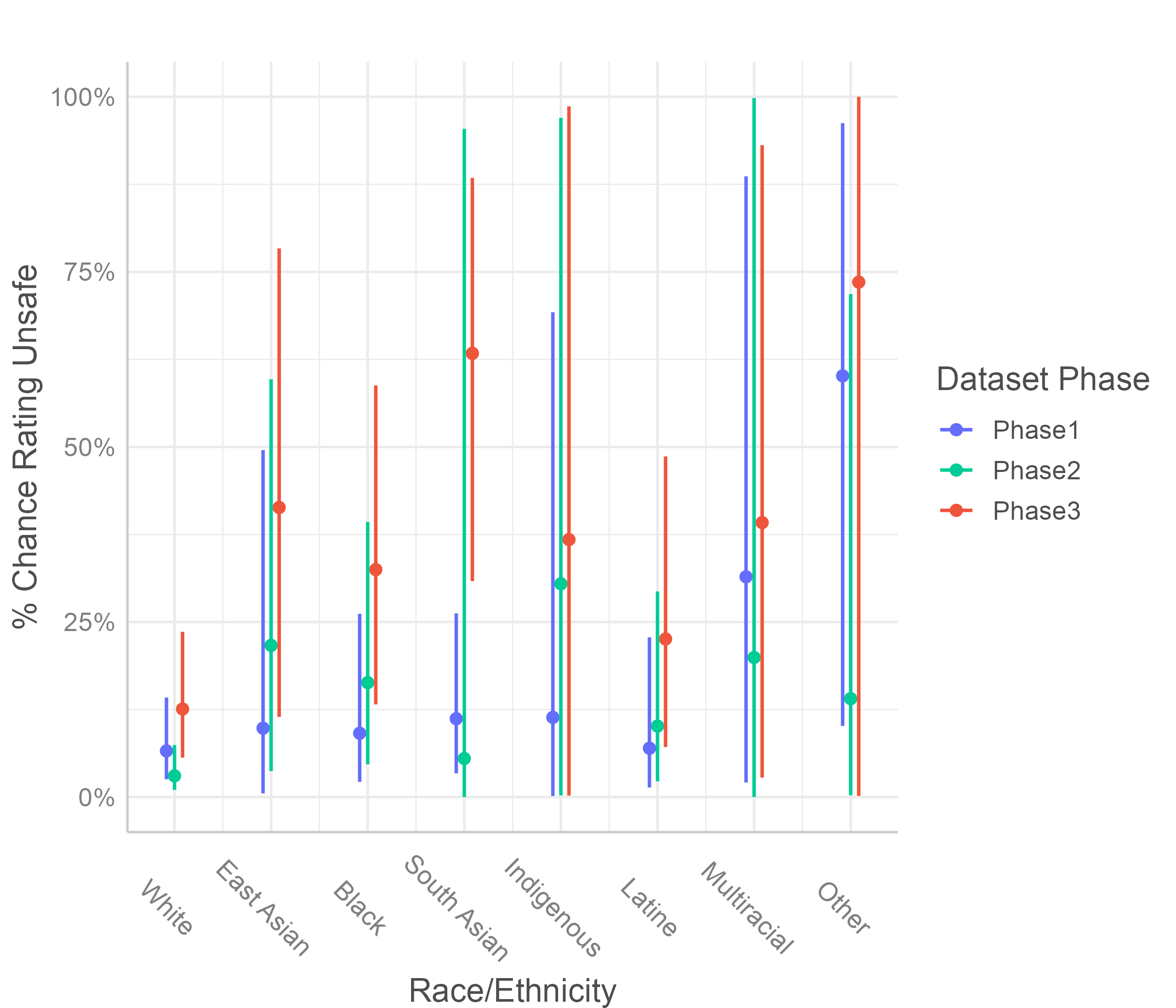}
    \caption{Plot of conditional effects of race/ethnicity and gender in the AD intersectional model (defined in Section \ref{sec:methods}) show that differences in safety ratings between phases 1 and 3 are large for raters of Color, compared to those for White raters. Non-White raters show greater variability in safety reporting across different conversational content. Plot holds rater age and education constant at their average values.}
    \label{fig:race_phase}
\end{figure}

Notably, dataset phase relates to the largest differences in safety ratings between raters, particularly when looking at the effects of phase across racial identities. These differences are sharpest between phases 1 and 3 (Figure \ref{fig:race_phase}). Compared to changes seen for White raters, East Asian, Black, Latine, Indigenous and especially South Asian raters show a heightened sensitivity to Phase 3 (highlighted in red) conversations compared to Phase 1 conversations (highlighted in blue). In particular, compared to phase 1, South Asian raters are nearly six times less likely to report a phase 3 conversation safe (63\% vs. 11\%). In contrast, White raters show only a modest jump, from a 7\% chance to a mere 13\% chance of reporting conversations as unsafe in Phases 1 and 3, respectively. Indeed, compared to White raters, South Asian raters are 3.31 times more likely to find Phase 3 content more unsafe (cf. Phase 1 content)---this difference has a 100\% chance of being positive, practically significant, and quite large (see Table \ref{tab:ad_sexit}).

Given the large differences in safety ratings across race and data collection phase, we form the exploratory hypothesis that \textit{raters of color have relatively higher sensitivity to unsafe content compared to White raters}. However, this hypothesis relies on the assumption that data collection phase is a proxy indicator of conversation severity, given the knowledge that, by design, Phase 3 conversations have a higher proportion of gold-labeled unsafe content than the other phases. Further, we form this hypothesis knowing that these differences are not due to data imbalances, as MLMs statistically account for imbalanced data.

\subsection{Results from QS models}
\label{results-qs}
Here we discuss results of the QS intersectional model, described in Section \ref{sec:methods}, zeroing in on how the effects of racial identity on safety ratings vary according to the severity of the content being rated. This additional model facilitates a more robust interrogation of the claim that raters of color are more sensitive to conversation severity than White raters. As discussed in Section \ref{sec:methods}, in this model, we subset data from dataset phases 1, 2, and 3 that had expert severity ratings, as shown in Table \ref{tab:data_harm_severity} in the supplemental materials. Since the severity variable is collinear with dataset phase, we eliminate \textit{phase} as a covariate in these remaining analyses.

\subsubsection{Strong independent AND intersectional effects for content severity.}
Across all demographic characteristics, conversation degree of harm is a highly reliable predictor of rating content as \textit{unsafe} (pd = 1.00; ps = 1.00) and is 90\% likely to be a large ($>$ $|.30|$) effect (see Table \ref{tab:qs_sexit}, row 12). But does the effect of racial/ethnic identity on safety rating depend on conversation's degree of harm (e.g., when a conversation's degree of harm is extremely unsafe)? We find this hypothesized effect is only certain and meaningful for Indigenous raters. 
As shown in Figure \ref{fig:QS_race_severity}, the chances of an Indigenous person finding a conversation \textit{unsafe} jumps from 20\% to 53\% when the degree of harm rating is \textit{benign} and \textit{extreme}, respectively. In other words, we estimate that Indigenous raters are 1.97 ($1.254^3$) times more likely than White raters to find a conversation more unsafe for conversations deemed by experts to be extremely unsafe (compared to safe; see Table \ref{tab:summary-qs-models}). Though only 23\% of this effect's posterior distribution is considered large ($>$ $.30$), this effect has a 99\% probability of being positive and a 96\% probability of being non-negligible (see Table \ref{tab:qs_sexit}, line 41 in the supplemental materials).

\begin{figure}
    \centering
    \includegraphics[width=.45\textwidth]{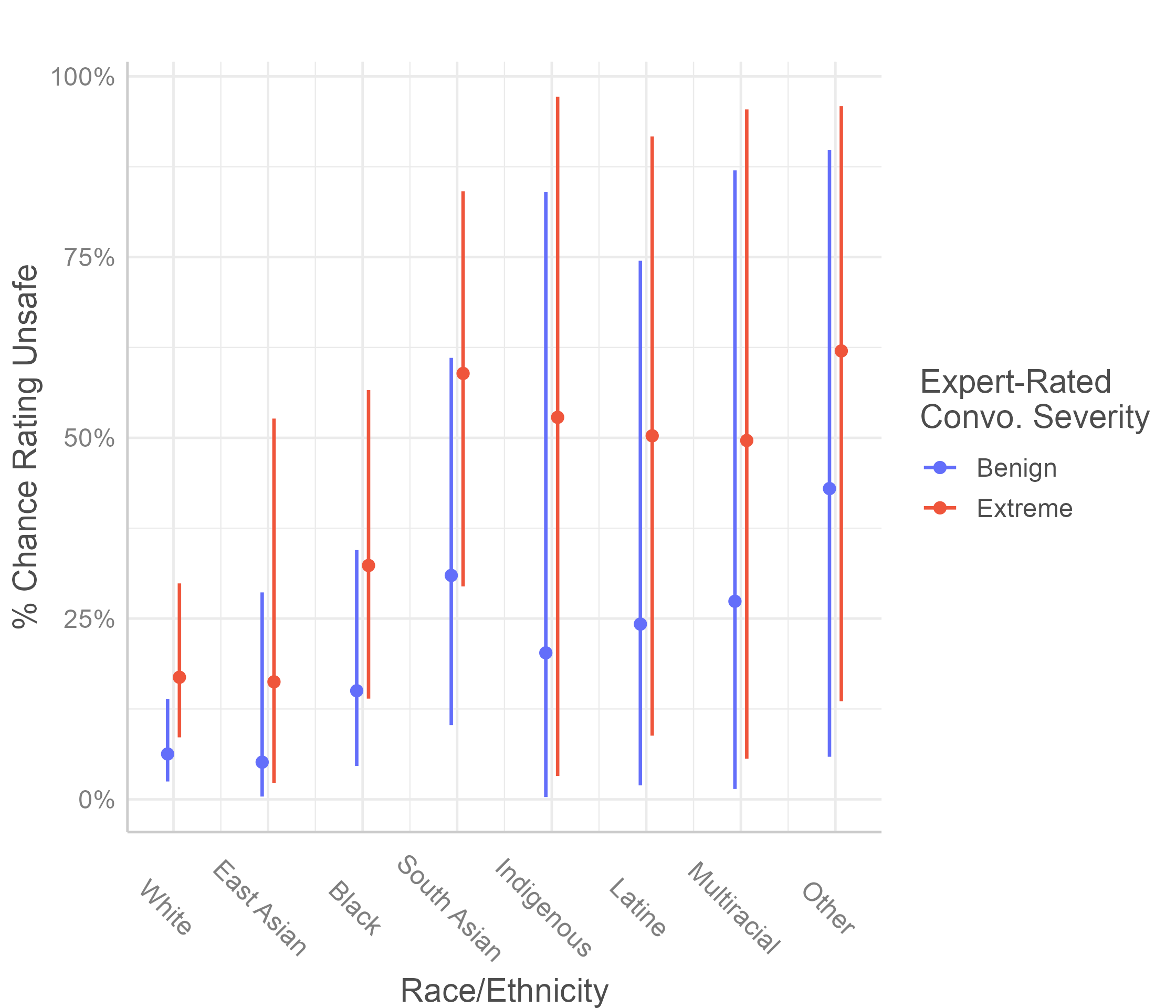}
    \caption{Conditional effects for QS intersectional Bayesian MLM (Section \ref{sec:methods}) show that South Asian and (especially) Indigenous raters are more sensitive to safety risks than raters of other racial/ethnicity groups when expert-rated degree of harm is high. The chances of Indigenous raters reporting \textit{unsafe} increases 33\% (from 20\% to 53\%) for ``Extreme" rather than ``Benign" conversations. Meanwhile, White raters' \textit{unsafe} ratings for these two content types only differ by 11\%.  We plot women as the reference group and hold age and education constant at their average values.}
    \label{fig:QS_race_severity}
\end{figure}

Meanwhile, harmful content likely relates to (slightly) heightened \textit{unsafe} ratings from South Asian and Latine raters, compared to White raters (pd = .91 and .88, respectively). It is probable, but not certain, that these rater groups’ safety ratings are more influenced by "Extreme" content compared to White raters. Probabilities that these effects are practically significant are questionable (ps = .72 and .77, for South Asian and Latine raters, respectively; see Table \ref{tab:qs_sexit}, lines 40 and 42 in the supplemental materials). 

\subsection{Results from both AD and QS models}
\subsubsection{Education level impacts safety ratings for Indigenous raters, but not other racial/ethnic groups.}
A striking result of both our final AD and QS models is that rater education levels are largely unrelated to safety reports across most demographic groups, but they are clearly linked to Indigenous raters' reports of safety. Indigenous raters, compared to White raters, are a staggering 3.12 times more likely (95\% Bayesian CI = [0.79, 15.71]) to report content as unsafe in QS Model 3.1, but only when their level of education is at the high school level or below. Holding all other factors constant, this effect is 94\% likely to exist, 94\% likely to be non-negligible, and 88\% likely to be large (see Table \ref{tab:qs_sexit}).





\section{Discussion}
\label{sec:discussion}
Our experiments with Bayesian multi-level modeling suggest that demographics do play a powerful role in predicting rater perceptions of safety in evaluation of conversational AI systems. Our intersectional models had roughly the same predictive power as our linear models. However, the intersectional models provide a nuanced view at how predictors interact, which is critical for their in-depth understanding. 

The results show \textit{strong intersectional effects involving race/ethnicity} that do not exist for race/ethnicity independently. That is, the effects of race/ethnicity on safety ratings \textit{only} emerge when race/ethnicity is viewed at its intersection with additional factors, like gender or harm severity of the conversation. In particular, South Asian women are more likely, and East Asian women less likely, than White raters to report conversations as \emph{Unsafe}. Indigenous, South Asian, and Latine raters are more likely than White raters to report conversations as \emph{Unsafe}. On the other hand, \textit{age is a strong independent predictor of rating behavior}, with younger raters more likely to rate conversations \emph{Unsafe}.

\begin{figure}
    \centering
    \includegraphics[width=.45\textwidth]{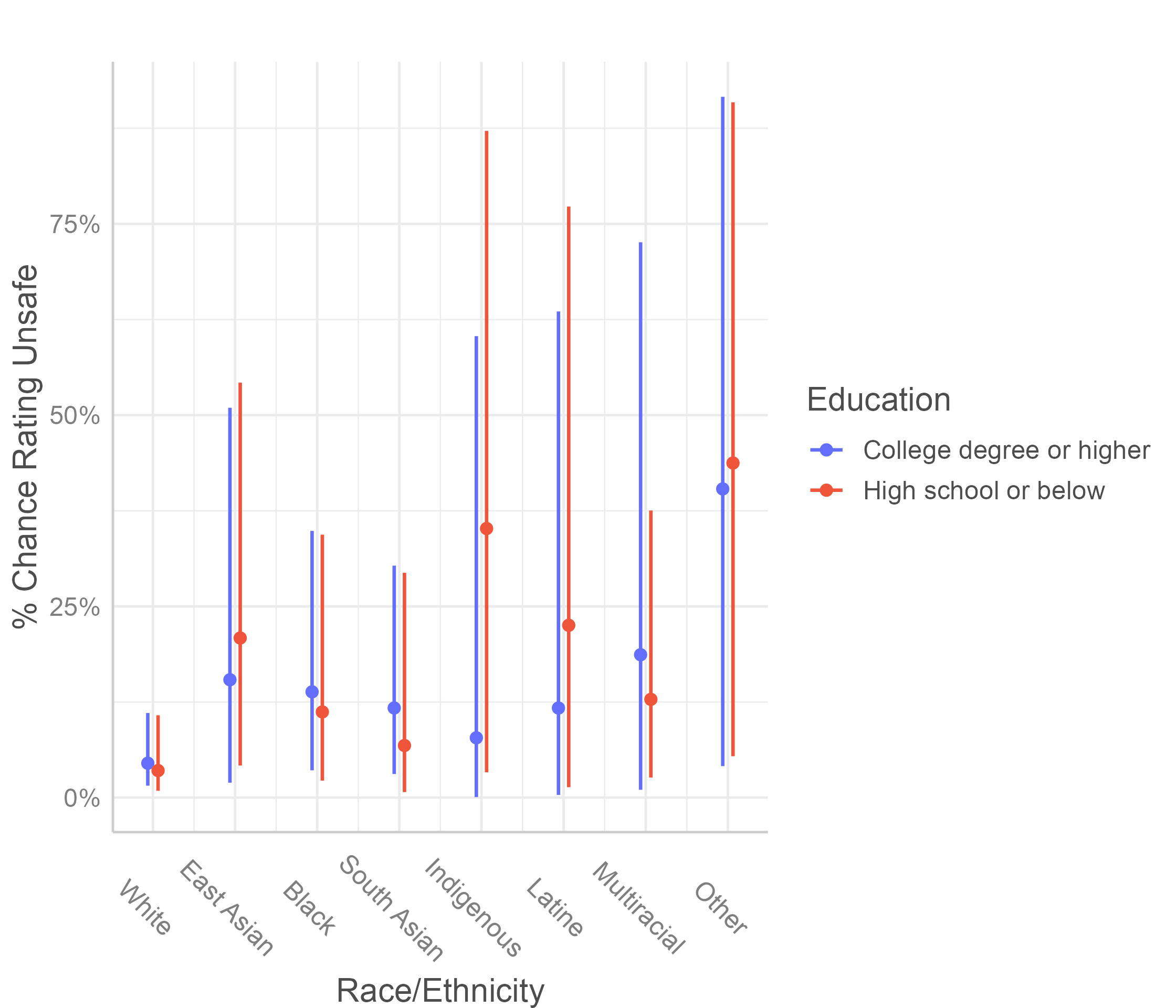}
    \caption{Conditional effects plot of the final QS model shows that race/ethnicity and education intersect for Indigenous raters with a high school level or below of education, even when holding age and gender constant at "Millenial" and "Man."}
    \label{fig:QS_race_education}
\end{figure}

Hence, we recommend that safety evaluation workflows recruit human raters across a broad demographic spectrum and record the demographic characteristics of raters to ensure that such breadth is maintained. To boost the representational power of demographic diversity, large rater pools should be used, considering the benefits that such diversity provides in weighing costs. In cases where costs are prohibitive, decreasing the number of items each rater evaluates should be considered in favor of increased number of raters per item. Finally, we recommend using statistical frameworks that account for the cross-classified structure of human annotation data.

\subsection{Limitations}
Although Bayesian MLMs depend on far fewer assumptions than linear regression or ANOVAs, there are some drawbacks. MCMC sampling is a slow process; our largest models take days to run if not parallelized across multiple CPUs, and it is relatively common for the process not to converge. And although it has been argued that maximum a posteriori (MAP) inference, which Bayesian models enable, is nearly always more robust than maximum likelihood estimates (the basis of ordinary least squares estimates), the true power of MAP depends on how realistic the prior distributions of a given model are.  

Although our models predict a unique intercept for each rater\_id and conversation\_id, the contribution from each rater and conversation pair is linear. We did not explore whether the relationship between them was more complex.



In this study, we only considered safety ratings as a single 
response (i.e. Q\_overall) for each (conversation, rater) pair. However, this response is an aggregate of 16--24 safety-related questions (i.e., safety dimensions discussed in \S \ref{sec:data}). In future work, the approach introduced by CrowdTruth \cite{aroyo2015truth} where raters, content, and questions are assumed to be dependent, could allow us to model the responses to these individual safety dimensions as a random effect.

Our \emph{Indigenous} race/ethnicity category lumps together very diverse Indigenous identities in a manner that likely discounts rich idiographic differences in language, culture, and lived experience \cite{else2016intersectionality}. However, in the interest of protecting participants privacy and prioritizing the representation of Indigenous perspectives in this empirical research, we chose to group them together. Creating the \emph{Indigenous} category in our analysis balances these opposing concerns, but leaves significant room for future study.

\section{Conclusion}
\label{sec:conclusion}
We apply Bayesian multilevel models (MLMs) to a dataset of 1,340 chatbot conversations, each rated for safety by 60--104 human raters, to study the impact of rater demographics on rater behavior for safety ratings. MLMs allow us to deal with the overlapping hierarchical dependencies on rater and conversation that are inherent in rater data, and which confound simpler modeling approaches, such as ordinary least squares regression and ANOVA. 

Our results show strong intersectional effects between race/ethnicity and gender, Indigenous raters and education, and content severity and race. They suggest that conversational AI safety evaluation can benefit when human evaluators come from diverse demographic backgrounds.

 \bibliography{references}

\clearpage

\section{Appendix: Full tables}



\begin{table*}
\tiny
\centering
\begin{tabular}[width=1\textwidth]{lddddd }
\toprule
  & {QS Null} & {QS Effects} & {QS Effects GE} & {QS Intersectional} & {QS Intersectional GE}\\
\midrule
Intercept[1] & {1.470 [1.305, 1.669]} & {2.415 [1.797, 3.244]} & {2.428 [1.807, 3.234]} & {2.187 [1.571, 3.061]} & {2.140 [1.531, 2.958]}\\
Intercept[2] & {1.866 [1.656, 2.118]} & {3.110 [2.313, 4.175]} & {3.131 [2.326, 4.171]} & {2.816 [2.022, 3.954]} & {2.760 [1.978, 3.812]}\\
RaceEastAsian & {} & {0.590 [0.401, 0.850]} & {0.636 [0.439, 0.923]} & {1.340 [0.730, 2.512]} & {1.269 [0.701, 2.367]}\\
RaceBlack & {} & {1.204 [0.804, 1.781]} & {1.294 [0.879, 1.875]} & {1.210 [0.685, 2.185]} & {1.219 [0.697, 2.108]}\\
RaceSouthAsian & {} & {3.113 [2.228, 4.359]} & {3.157 [2.279, 4.524]} & {1.793 [1.113, 2.862]} & {1.742 [1.075, 2.796]}\\
RaceIndigenous & {} & {2.437 [1.307, 4.682]} & {2.494 [1.300, 4.669]} & {1.572 [0.564, 4.671]} & {1.454 [0.502, 4.319]}\\
RaceLatine & {} & {1.452 [0.749, 2.747]} & {1.404 [0.738, 2.662]} & {1.188 [0.328, 4.264]} & {1.243 [0.352, 4.781]}\\
RaceMultiracial & {} & {1.386 [0.796, 2.427]} & {1.453 [0.849, 2.464]} & {1.510 [0.572, 4.158]} & {1.531 [0.599, 3.977]}\\
RaceOther & {} & {1.591 [0.785, 3.175]} & {1.685 [0.874, 3.357]} & {1.607 [0.672, 3.913]} & {1.595 [0.660, 3.947]}\\
GenderWoman & {} & {1.247 [0.959, 1.628]} & {1.230 [0.954, 1.582]} & {1.210 [0.797, 1.847]} & {1.178 [0.789, 1.783]}\\
Age.L & {} & {0.723 [0.638, 0.819]} & {0.734 [0.646, 0.835]} & {0.554 [0.439, 0.694]} & {0.550 [0.439, 0.689]}\\
Severity.L & {} & {1.575 [1.402, 1.777]} & {1.592 [1.396, 1.813]} & {1.478 [1.288, 1.688]} & {1.490 [1.284, 1.741]}\\
EducationHighschoolorbelow & {} & {1.219 [0.889, 1.699]} & {1.108 [0.803, 1.559]} & {0.939 [0.587, 1.531]} & {0.890 [0.553, 1.433]}\\
EducationOther & {} & {0.461 [0.282, 0.750]} & {0.448 [0.271, 0.732]} & {0.812 [0.298, 2.161]} & {0.813 [0.292, 2.159]}\\
RaceEastAsian:GenderWoman & {} & {} & {} & {0.437 [0.211, 0.869]} & {0.461 [0.223, 0.910]}\\
RaceBlack:GenderWoman & {} & {} & {} & {0.854 [0.407, 1.803]} & {0.895 [0.433, 1.826]}\\
RaceSouthAsian:GenderWoman & {} & {} & {} & {1.679 [0.924, 3.159]} & {1.690 [0.936, 3.157]}\\
RaceIndigenous:GenderWoman & {} & {} & {} & {1.372 [0.439, 4.632]} & {1.521 [0.488, 5.221]}\\
RaceLatine:GenderWoman & {} & {} & {} & {1.422 [0.437, 4.834]} & {1.372 [0.431, 4.648]}\\
RaceMultiracial:GenderWoman & {} & {} & {} & {1.026 [0.360, 2.892]} & {1.133 [0.416, 3.205]}\\
RaceOther:GenderWoman & {} & {} & {} & {0.923 [0.272, 3.002]} & {0.899 [0.271, 2.913]}\\
RaceEastAsian:Age.L & {} & {} & {} & {1.591 [1.200, 2.117]} & {1.655 [1.251, 2.206]}\\
RaceBlack:Age.L & {} & {} & {} & {1.501 [0.803, 2.856]} & {1.515 [0.836, 2.807]}\\
RaceSouthAsian:Age.L & {} & {} & {} & {1.181 [0.774, 1.804]} & {1.239 [0.802, 1.873]}\\
RaceIndigenous:Age.L & {} & {} & {} & {1.161 [0.291, 4.816]} & {1.261 [0.320, 5.612]}\\
RaceLatine:Age.L & {} & {} & {} & {0.932 [0.141, 6.425]} & {0.891 [0.118, 6.195]}\\
RaceMultiracial:Age.L & {} & {} & {} & {1.256 [0.459, 3.609]} & {1.244 [0.464, 3.425]}\\
RaceOther:Age.L & {} & {} & {} & {1.168 [0.363, 4.059]} & {1.191 [0.368, 4.155]}\\
RaceEastAsian:Severity.L & {} & {} & {} & {1.028 [0.918, 1.151]} & {1.040 [0.902, 1.201]}\\
RaceBlack:Severity.L & {} & {} & {} & {0.986 [0.893, 1.090]} & {0.986 [0.868, 1.119]}\\
RaceSouthAsian:Severity.L & {} & {} & {} & {1.074 [0.967, 1.193]} & {1.092 [0.959, 1.241]}\\
RaceIndigenous:Severity.L & {} & {} & {} & {1.258 [1.076, 1.476]} & {1.254 [1.028, 1.534]}\\
RaceLatine:Severity.L & {} & {} & {} & {1.119 [0.919, 1.365]} & {1.147 [0.904, 1.462]}\\
RaceMultiracial:Severity.L & {} & {} & {} & {1.001 [0.871, 1.153]} & {1.008 [0.846, 1.201]}\\
RaceOther:Severity.L & {} & {} & {} & {0.908 [0.720, 1.135]} & {0.890 [0.677, 1.173]}\\
RaceEastAsian:EducationHighschoolorbelow & {} & {} & {} & {1.367 [0.522, 3.608]} & {1.365 [0.543, 3.501]}\\
RaceBlack:EducationHighschoolorbelow & {} & {} & {} & {0.991 [0.440, 2.197]} & {0.992 [0.449, 2.140]}\\
RaceSouthAsian:EducationHighschoolorbelow & {} & {} & {} & {0.804 [0.326, 1.925]} & {0.839 [0.347, 1.982]}\\
RaceIndigenous:EducationHighschoolorbelow & {} & {} & {} & {2.943 [0.770, 15.355]} & {3.115 [0.794, 15.705]}\\
RaceLatine:EducationHighschoolorbelow & {} & {} & {} & {1.401 [0.298, 8.905]} & {1.661 [0.351, 10.717]}\\
RaceMultiracial:EducationHighschoolorbelow & {} & {} & {} & {0.880 [0.214, 3.297]} & {0.876 [0.224, 3.255]}\\
RaceOther:EducationHighschoolorbelow & {} & {} & {} & {1.130 [0.225, 5.821]} & {1.181 [0.239, 6.979]}\\
RaceEastAsian:EducationOther & {} & {} & {} & {1.009 [0.043, 24.408]} & {1.005 [0.046, 24.839]}\\
RaceBlack:EducationOther & {} & {} & {} & {1.126 [0.319, 3.875]} & {0.973 [0.283, 3.368]}\\
RaceSouthAsian:EducationOther & {} & {} & {} & {1.808 [0.534, 7.021]} & {1.767 [0.527, 6.878]}\\
RaceIndigenous:EducationOther & {} & {} & {} & {1.043 [0.164, 6.863]} & {0.929 [0.136, 5.815]}\\
RaceLatine:EducationOther & {} & {} & {} & {0.995 [0.038, 21.825]} & {0.976 [0.022, 22.225]}\\
RaceMultiracial:EducationOther & {} & {} & {} & {1.001 [0.045, 21.327]} & {0.997 [0.041, 27.361]}\\
RaceOther:EducationOther & {} & {} & {} & {0.998 [0.036, 26.575]} & {0.998 [0.038, 27.866]}\\
\midrule
sd\_dialogue\_id\_\_Intercept & {2.215 [2.116, 2.323]} & {2.064 [1.984, 2.161]} & {2.073 [1.989, 2.173]} & {2.065 [1.979, 2.160]} & {2.075 [1.991, 2.171]}\\
sd\_rater\_id\_\_Intercept & {2.302 [2.140, 2.511]} & {2.429 [2.201, 2.728]} & {2.402 [2.179, 2.701]} & {2.349 [2.133, 2.634]} & {2.351 [2.133, 2.651]}\\
sd\_rater\_id\_\_Severity.L & {} & {} & {1.170 [1.111, 1.235]} & {} & {1.165 [1.105, 1.231]}\\
cor\_rater\_id\_\_Intercept\_\_Severity.L & {} & {} & {1.013 [0.742, 1.384]} & {} & {0.952 [0.682, 1.337]}\\
\bottomrule
\end{tabular}
\caption{Summary of Results for MLMs with Expert-Annotated Data. Point estimates are presented as odds ratios, with 95\% credible intervals provided in brackets beside each estimate. Population-level effects (i.e., effects concerning all rating observations across raters and conversations) are presented first, followed by the structure of the group-level effects (i.e., effects concerning the grouping variables of \textit{rater\_id} and \textit{conversation\_id}), which are shown at the bottom of the table.}
\label{tab:summary-qs-models}
\end{table*}

\begin{table*}
\tiny
\centering
\begin{tabular}[width=1\textwidth]{lddd}
\toprule
  & {AD Null} & {AD Effects} & {AD Intersectional}\\
\midrule
Intercept[1] & {2.053 [1.845, 2.304]} & {3.739 [2.755, 5.065]} & {2.990 [2.180, 4.123]}\\
Intercept[2] & {2.615 [2.348, 2.934]} & {4.819 [3.551, 6.525]} & {3.856 [2.811, 5.318]}\\
RaceEastAsian & {} & {0.529 [0.364, 0.765]} & {0.874 [0.358, 2.072]}\\
RaceBlack & {} & {1.141 [0.783, 1.700]} & {0.825 [0.461, 1.483]}\\
RaceSouthAsian & {} & {3.077 [2.174, 4.389]} & {1.274 [0.815, 2.000]}\\
RaceIndigenous & {} & {2.272 [1.235, 4.229]} & {1.178 [0.479, 2.975]}\\
RaceLatine & {} & {1.283 [0.886, 1.859]} & {0.806 [0.452, 1.438]}\\
RaceMultiracial & {} & {1.328 [0.763, 2.371]} & {1.688 [0.521, 6.391]}\\
RaceOther & {} & {1.666 [0.844, 3.296]} & {2.599 [0.890, 8.277]}\\
GenderWoman & {} & {1.250 [0.977, 1.590]} & {1.274 [0.853, 1.881]}\\
Age.L & {} & {0.666 [0.606, 0.736]} & {0.651 [0.547, 0.770]}\\
EducationHighschoolorbelow & {} & {1.313 [0.950, 1.783]} & {0.980 [0.627, 1.531]}\\
EducationOther & {} & {0.534 [0.350, 0.819]} & {0.646 [0.253, 1.589]}\\
Phase1vsPhase2 & {} & {0.849 [0.770, 0.935]} & {0.692 [0.604, 0.794]}\\
Phase1vsPhase3 & {} & {1.771 [1.531, 2.047]} & {1.432 [1.191, 1.732]}\\
RaceEastAsian:GenderWoman & {} & {} & {0.465 [0.232, 0.928]}\\
RaceBlack:GenderWoman & {} & {} & {0.752 [0.360, 1.569]}\\
RaceSouthAsian:GenderWoman & {} & {} & {1.564 [0.873, 2.826]}\\
RaceIndigenous:GenderWoman & {} & {} & {1.133 [0.338, 3.769]}\\
RaceLatine:GenderWoman & {} & {} & {0.968 [0.510, 1.873]}\\
RaceMultiracial:GenderWoman & {} & {} & {0.948 [0.338, 2.619]}\\
RaceOther:GenderWoman & {} & {} & {0.898 [0.285, 2.845]}\\
RaceEastAsian:Age.L & {} & {} & {1.259 [0.980, 1.633]}\\
RaceBlack:Age.L & {} & {} & {1.311 [0.739, 2.350]}\\
RaceSouthAsian:Age.L & {} & {} & {1.182 [0.776, 1.791]}\\
RaceIndigenous:Age.L & {} & {} & {1.178 [0.375, 3.837]}\\
RaceLatine:Age.L & {} & {} & {1.218 [0.715, 2.051]}\\
RaceMultiracial:Age.L & {} & {} & {0.961 [0.285, 3.133]}\\
RaceOther:Age.L & {} & {} & {0.970 [0.311, 3.020]}\\
RaceEastAsian:PhasePhase2 & {} & {} & {2.382 [1.118, 5.267]}\\
RaceBlack:PhasePhase2 & {} & {} & {2.060 [1.496, 2.829]}\\
RaceSouthAsian:PhasePhase2 & {} & {} & {0.999 [0.037, 24.697]}\\
RaceIndigenous:PhasePhase2 & {} & {} & {2.797 [0.672, 16.026]}\\
RaceLatine:PhasePhase2 & {} & {} & {1.773 [1.347, 2.341]}\\
RaceMultiracial:PhasePhase2 & {} & {} & {0.998 [0.035, 29.616]}\\
RaceOther:PhasePhase2 & {} & {} & {0.391 [0.102, 1.232]}\\
RaceEastAsian:PhasePhase3 & {} & {} & {2.023 [0.975, 4.399]}\\
RaceBlack:PhasePhase3 & {} & {} & {1.680 [1.301, 2.175]}\\
RaceSouthAsian:PhasePhase3 & {} & {} & {3.312 [1.738, 6.386]}\\
RaceIndigenous:PhasePhase3 & {} & {} & {1.630 [0.417, 7.748]}\\
RaceLatine:PhasePhase3 & {} & {} & {1.438 [1.087, 1.903]}\\
RaceMultiracial:PhasePhase3 & {} & {} & {0.877 [0.175, 3.765]}\\
RaceOther:PhasePhase3 & {} & {} & {1.000 [0.041, 26.662]}\\
RaceEastAsian:EducationHighschoolorbelow & {} & {} & {1.180 [0.473, 2.946]}\\
RaceBlack:EducationHighschoolorbelow & {} & {} & {1.048 [0.480, 2.225]}\\
RaceSouthAsian:EducationHighschoolorbelow & {} & {} & {0.958 [0.413, 2.256]}\\
RaceIndigenous:EducationHighschoolorbelow & {} & {} & {2.640 [0.646, 15.430]}\\
RaceLatine:EducationHighschoolorbelow & {} & {} & {1.790 [0.808, 4.036]}\\
RaceMultiracial:EducationHighschoolorbelow & {} & {} & {0.939 [0.181, 4.356]}\\
RaceOther:EducationHighschoolorbelow & {} & {} & {1.853 [0.376, 14.610]}\\
RaceEastAsian:EducationOther & {} & {} & {0.993 [0.048, 22.956]}\\
RaceBlack:EducationOther & {} & {} & {1.350 [0.419, 4.749]}\\
RaceSouthAsian:EducationOther & {} & {} & {0.917 [0.236, 3.364]}\\
RaceIndigenous:EducationOther & {} & {} & {1.277 [0.273, 6.928]}\\
RaceLatine:EducationOther & {} & {} & {0.789 [0.173, 3.140]}\\
RaceMultiracial:EducationOther & {} & {} & {1.000 [0.036, 23.332]}\\
RaceOther:EducationOther & {} & {} & {1.012 [0.045, 29.572]}\\
\midrule
sd\_conversation\_id\_\_Intercept & {2.439 [2.351, 2.534]} & {2.376 [2.295, 2.471]} & {2.379 [2.297, 2.469]}\\
sd\_rater\_id\_\_Intercept & {2.329 [2.169, 2.539]} & {2.463 [2.251, 2.739]} & {2.246 [2.066, 2.481]}\\
\bottomrule
\end{tabular}
\caption{Summary of results for AD MLMs with data collection phase as a variable, but no expert ratings. Point estimates are presented as odds ratios, with 95\% credible intervals provided in brackets beside each estimate. Population-level effects (i.e., effects concerning all rating observations across raters and conversations) are presented first, followed by the structure of the group-level effects (i.e., effects concerning the grouping variables of \textit{rater\_id} and \textit{conversation\_id}), which are shown at the bottom of the table.}
\label{tab:summary-main-models}
\end{table*}

\begin{table*}
\tiny
\centering
\begin{tabular}{llrlrrr}
\toprule
{} &                  Parameter &  Median &                  CI &  Direction & Significance &  Large \\
\midrule
0  &             b\_Intercept[1] &  0.7608 &    [0.4258, 1.0845] &     0.9998 &        0.9998 & 0.9956 \\
1  &             b\_Intercept[2] &  1.0151 &    [0.6821, 1.3381] &     1.0000 &        1.0000 & 0.9998 \\
2  &                b\_EastAsian &  0.2383 &   [-0.3550, 0.8617] &     0.7845 &        0.7308 & 0.4210 \\
3  &                    b\_Black &  0.1984 &   [-0.3613, 0.7458] &     0.7506 &        0.6965 & 0.3633 \\
4  &               b\_SouthAsian &  0.5550 &    [0.0724, 1.0283] &     0.9862 &        0.9792 & 0.8612 \\
5  &               b\_Indigenous &  0.3744 &   [-0.6894, 1.4631] &     0.7596 &        0.7262 & 0.5528 \\
6  &                  b\_Latinxe &  0.2179 &   [-1.0445, 1.5647] &     0.6353 &        0.6032 & 0.4457 \\
7  &              b\_Multiracial &  0.4259 &   [-0.5121, 1.3806] &     0.8133 &        0.7863 & 0.6048 \\
8  &                    b\_Other &  0.4671 &   [-0.4150, 1.3729] &     0.8458 &        0.8194 & 0.6436 \\
9  &                    b\_Woman &  0.1641 &   [-0.2376, 0.5781] &     0.7837 &        0.7083 & 0.2586 \\
10 &                    b\_Age.L & -0.5981 &  [-0.8227, -0.3722] &     1.0000 &        1.0000 & 0.9956 \\
11 &                    b\_Age.Q &  0.2357 &   [-0.1184, 0.5936] &     0.9033 &        0.8458 & 0.3655 \\
12 &               b\_Severity.L &  0.3986 &    [0.2498, 0.5546] &     1.0000 &        0.9999 & 0.8966 \\
13 &               b\_Severity.Q & -0.4039 &  [-0.5453, -0.2628] &     1.0000 &        1.0000 & 0.9265 \\
14 &               b\_Severity.C &  0.0823 &   [-0.0619, 0.2272] &     0.8684 &        0.6708 & 0.0017 \\
15 &              b\_Highschool- & -0.1169 &   [-0.5919, 0.3597] &     0.6870 &        0.6122 & 0.2212 \\
16 &                    b\_Other & -0.2072 &   [-1.2300, 0.7695] &     0.6564 &        0.6215 & 0.4281 \\
17 &          b\_EastAsian:Woman & -0.7747 &  [-1.5025, -0.0948] &     0.9870 &        0.9806 & 0.9135 \\
18 &              b\_Black:Woman & -0.1107 &   [-0.8372, 0.6021] &     0.6188 &        0.5673 & 0.2985 \\
19 &         b\_SouthAsian:Woman &  0.5249 &   [-0.0663, 1.1495] &     0.9587 &        0.9418 & 0.7685 \\
20 &         b\_Indigenous:Woman &  0.4196 &   [-0.7180, 1.6526] &     0.7636 &        0.7347 & 0.5798 \\
21 &            b\_Latinxe:Woman &  0.3161 &   [-0.8419, 1.5364] &     0.7070 &        0.6762 & 0.5098 \\
22 &        b\_Multiracial:Woman &  0.1250 &   [-0.8760, 1.1648] &     0.5973 &        0.5597 & 0.3645 \\
23 &              b\_Other:Woman & -0.1063 &   [-1.3059, 1.0692] &     0.5706 &        0.5375 & 0.3726 \\
24 &          b\_EastAsian:Age.L &  0.5039 &    [0.2243, 0.7911] &     0.9994 &        0.9989 & 0.9172 \\
25 &              b\_Black:Age.L &  0.4153 &   [-0.1789, 1.0320] &     0.9134 &        0.8834 & 0.6423 \\
26 &         b\_SouthAsian:Age.L &  0.2141 &   [-0.2209, 0.6275] &     0.8413 &        0.7801 & 0.3485 \\
27 &         b\_Indigenous:Age.L &  0.2317 &   [-1.1391, 1.7249] &     0.6382 &        0.6102 & 0.4601 \\
28 &            b\_Latinxe:Age.L & -0.1159 &   [-2.1407, 1.8238] &     0.5519 &        0.5317 & 0.4170 \\
29 &        b\_Multiracial:Age.L &  0.2186 &   [-0.7678, 1.2310] &     0.6694 &        0.6334 & 0.4356 \\
30 &              b\_Other:Age.L &  0.1747 &   [-1.0005, 1.4244] &     0.6144 &        0.5796 & 0.4185 \\
31 &          b\_EastAsian:Age.Q & -0.5459 &   [-1.4080, 0.2970] &     0.8966 &        0.8730 & 0.7188 \\
32 &              b\_Black:Age.Q & -0.4592 &   [-1.0752, 0.1535] &     0.9294 &        0.9048 & 0.6955 \\
33 &         b\_SouthAsian:Age.Q &  0.0194 &   [-0.6468, 0.6718] &     0.5233 &        0.4598 & 0.1927 \\
34 &         b\_Indigenous:Age.Q & -0.0941 &   [-1.5662, 1.3473] &     0.5537 &        0.5246 & 0.3876 \\
35 &            b\_Latinxe:Age.Q & -0.3316 &   [-1.8472, 1.0426] &     0.6855 &        0.6600 & 0.5187 \\
36 &        b\_Multiracial:Age.Q & -0.4605 &   [-1.6888, 0.6533] &     0.7909 &        0.7623 & 0.6122 \\
37 &              b\_Other:Age.Q & -1.1040 &   [-2.5653, 0.1794] &     0.9539 &        0.9452 & 0.8872 \\
38 &     b\_EastAsian:Severity.L &  0.0389 &   [-0.1036, 0.1831] &     0.7026 &        0.4392 & 0.0002 \\
39 &         b\_Black:Severity.L & -0.0143 &   [-0.1418, 0.1123] &     0.5861 &        0.2966 & 0.0000 \\
40 &    b\_SouthAsian:Severity.L &  0.0878 &   [-0.0414, 0.2161] &     0.9093 &        0.7198 & 0.0007 \\
41 &    b\_Indigenous:Severity.L &  0.2266 &    [0.0274, 0.4279] &     0.9877 &        0.9594 & 0.2333 \\
42 &       b\_Latinxe:Severity.L &  0.1368 &   [-0.1009, 0.3798] &     0.8761 &        0.7705 & 0.0953 \\
43 &   b\_Multiracial:Severity.L &  0.0082 &   [-0.1669, 0.1834] &     0.5383 &        0.3145 & 0.0004 \\
44 &         b\_Other:Severity.L & -0.1161 &   [-0.3903, 0.1599] &     0.7943 &        0.6777 & 0.0912 \\
45 &     b\_EastAsian:Severity.Q &  0.0565 &   [-0.0785, 0.1925] &     0.7987 &        0.5349 & 0.0000 \\
46 &         b\_Black:Severity.Q &  0.0696 &   [-0.0531, 0.1866] &     0.8668 &        0.6262 & 0.0002 \\
47 &    b\_SouthAsian:Severity.Q &  0.0656 &   [-0.0591, 0.1861] &     0.8495 &        0.6002 & 0.0001 \\
48 &    b\_Indigenous:Severity.Q & -0.0688 &   [-0.2581, 0.1185] &     0.7652 &        0.5796 & 0.0067 \\
49 &       b\_Latinxe:Severity.Q &  0.1175 &   [-0.1166, 0.3517] &     0.8393 &        0.7201 & 0.0634 \\
50 &   b\_Multiracial:Severity.Q &  0.0089 &   [-0.1614, 0.1840] &     0.5433 &        0.3152 & 0.0004 \\
51 &         b\_Other:Severity.Q &  0.0649 &   [-0.2062, 0.3276] &     0.6832 &        0.5430 & 0.0418 \\
52 &     b\_EastAsian:Severity.C &  0.0553 &   [-0.0722, 0.1816] &     0.8042 &        0.5361 & 0.0000 \\
53 &         b\_Black:Severity.C &  0.0557 &   [-0.0568, 0.1649] &     0.8353 &        0.5387 & 0.0001 \\
54 &    b\_SouthAsian:Severity.C &  0.0750 &   [-0.0418, 0.1913] &     0.9011 &        0.6633 & 0.0000 \\
55 &    b\_Indigenous:Severity.C &  0.0356 &   [-0.1353, 0.2132] &     0.6587 &        0.4345 & 0.0019 \\
56 &       b\_Latinxe:Severity.C & -0.1110 &   [-0.3372, 0.1149] &     0.8269 &        0.6976 & 0.0513 \\
57 &   b\_Multiracial:Severity.C &  0.0181 &   [-0.1416, 0.1808] &     0.5905 &        0.3462 & 0.0006 \\
58 &         b\_Other:Severity.C &  0.1535 &   [-0.1008, 0.4154] &     0.8842 &        0.7905 & 0.1334 \\
59 &    b\_EastAsian:Highschool- &  0.3109 &   [-0.6107, 1.2531] &     0.7543 &        0.7163 & 0.5102 \\
60 &        b\_Black:Highschool- & -0.0081 &   [-0.8004, 0.7606] &     0.5082 &        0.4577 & 0.2255 \\
61 &   b\_SouthAsian:Highschool- & -0.1750 &   [-1.0581, 0.6841] &     0.6582 &        0.6158 & 0.3846 \\
62 &   b\_Indigenous:Highschool- &  1.1361 &   [-0.2305, 2.7540] &     0.9433 &        0.9353 & 0.8779 \\
63 &      b\_Latinxe:Highschool- &  0.5075 &   [-1.0476, 2.3719] &     0.7349 &        0.7124 & 0.5992 \\
64 &  b\_Multiracial:Highschool- & -0.1320 &   [-1.4966, 1.1803] &     0.5776 &        0.5478 & 0.3980 \\
65 &        b\_Other:Highschool- &  0.1666 &   [-1.4311, 1.9430] &     0.5891 &        0.5622 & 0.4316 \\
66 &          b\_EastAsian:Other &  0.0046 &   [-3.0684, 3.2124] &     0.5016 &        0.4825 & 0.3927 \\
67 &              b\_Black:Other & -0.0273 &   [-1.2630, 1.2142] &     0.5180 &        0.4853 & 0.3264 \\
68 &         b\_SouthAsian:Other &  0.5695 &   [-0.6401, 1.9284] &     0.8208 &        0.7968 & 0.6653 \\
69 &         b\_Indigenous:Other & -0.0737 &   [-1.9951, 1.7605] &     0.5373 &        0.5130 & 0.3927 \\
70 &            b\_Latinxe:Other & -0.0240 &   [-3.8223, 3.1012] &     0.5091 &        0.4905 & 0.4043 \\
71 &        b\_Multiracial:Other & -0.0030 &   [-3.1947, 3.3091] &     0.5008 &        0.4836 & 0.3916 \\
72 &              b\_Other:Other & -0.0016 &   [-3.2720, 3.3274] &     0.5005 &        0.4818 & 0.3943 \\
\bottomrule
\end{tabular}
\caption{Summary of results for the \emph{AD intersectional} model with data collection phase as a variable, but no expert ratings. These results show standardized estimates according to the model for the class in question versus the variable's reference class that the conversation being rated \emph{safe}. Note that, for gender \emph{Man} is the reference class, for race/ethnicity it is \emph{White}, for dataset phase it is \emph{Phase 1}, and for education it is \emph{College degree or above}. The suffix \textit{.L} represents the estimated linear effect of a variable (e.g., from Gen Z to Millenial); \textit{.Q} represents the estimated quadratic effect of a variable (e.g., from Gen Z to Gen X+); \textit{.C} represents the estimated cubic effect of a variable (e.g., from conversation severity of \textit{Benign} to \textit{Extreme}.  \textit{Direction}, \textit{Significance}, and \textit{Large} correspond to probabilities of an effect's direction, practical significance, and magnitude being large, respectively.}
\label{tab:phase}
\label{tab:qs_sexit}
\end{table*}

\begin{table*}
\tiny
\centering
\begin{tabular}{llrlrrr}
\toprule
{} &                  Parameter &  Median &                  CI &  Direction &  Significance &  Larges \\
\midrule
0  &             b\_Intercept[1] &  1.0951 &    [0.7794, 1.4166] &     1.0000 &        1.0000 & 1.0000 \\
1  &             b\_Intercept[2] &  1.3497 &    [1.0335, 1.6711] &     1.0000 &        1.0000 & 1.0000 \\
2  &                b\_EastAsian & -0.1350 &   [-1.0263, 0.7287] &     0.6183 &        0.5743 & 0.3533 \\
3  &                    b\_Black & -0.1921 &   [-0.7741, 0.3942] &     0.7499 &        0.6884 & 0.3523 \\
4  &               b\_SouthAsian &  0.2425 &   [-0.2043, 0.6932] &     0.8619 &        0.8087 & 0.4002 \\
5  &               b\_Indigenous &  0.1642 &   [-0.7351, 1.0903] &     0.6417 &        0.6033 & 0.3828 \\
6  &                   b\_Latine & -0.2153 &   [-0.7945, 0.3633] &     0.7704 &        0.7153 & 0.3818 \\
7  &              b\_Multiracial &  0.5235 &   [-0.6525, 1.8548] &     0.8031 &        0.7788 & 0.6368 \\
8  &                    b\_Other &  0.9553 &   [-0.1170, 2.1135] &     0.9598 &        0.9525 & 0.8864 \\
9  &                    b\_Woman &  0.2420 &   [-0.1594, 0.6318] &     0.8861 &        0.8343 & 0.3793 \\
10 &                    b\_Age.L & -0.4296 &  [-0.6030, -0.2614] &     1.0000 &        1.0000 & 0.9357 \\
11 &                    b\_Age.Q &  0.1914 &   [-0.1495, 0.5467] &     0.8652 &        0.7960 & 0.2732 \\
12 &              b\_PhasePhase2 & -0.3677 &  [-0.5035, -0.2312] &     1.0000 &        1.0000 & 0.8333 \\
13 &              b\_PhasePhase3 &  0.3591 &    [0.1746, 0.5492] &     1.0000 &        0.9997 & 0.7334 \\
14 &              b\_Highschool- & -0.0202 &   [-0.4673, 0.4258] &     0.5376 &        0.4483 & 0.1062 \\
15 &                    b\_Other & -0.4376 &   [-1.3732, 0.4629] &     0.8284 &        0.7996 & 0.6119 \\
16 &          b\_EastAsian:Woman & -0.7662 &  [-1.4618, -0.0753] &     0.9861 &        0.9798 & 0.9064 \\
17 &              b\_Black:Woman & -0.2853 &   [-1.0210, 0.4502] &     0.7804 &        0.7414 & 0.4842 \\
18 &         b\_SouthAsian:Woman &  0.4475 &   [-0.1362, 1.0388] &     0.9357 &        0.9113 & 0.6902 \\
19 &         b\_Indigenous:Woman &  0.1252 &   [-1.0833, 1.3268] &     0.5838 &        0.5523 & 0.3807 \\
20 &             b\_Latine:Woman & -0.0329 &   [-0.6725, 0.6276] &     0.5417 &        0.4798 & 0.2077 \\
21 &        b\_Multiracial:Woman & -0.0530 &   [-1.0834, 0.9629] &     0.5413 &        0.5025 & 0.3150 \\
22 &              b\_Other:Woman & -0.1076 &   [-1.2542, 1.0454] &     0.5772 &        0.5415 & 0.3715 \\
23 &          b\_EastAsian:Age.L &  0.2306 &   [-0.0205, 0.4906] &     0.9641 &        0.9191 & 0.3000 \\
24 &              b\_Black:Age.L &  0.2711 &   [-0.3028, 0.8545] &     0.8220 &        0.7717 & 0.4610 \\
25 &         b\_SouthAsian:Age.L &  0.1671 &   [-0.2541, 0.5826] &     0.7832 &        0.7060 & 0.2706 \\
26 &         b\_Indigenous:Age.L &  0.1639 &   [-0.9803, 1.3448] &     0.6141 &        0.5812 & 0.4116 \\
27 &             b\_Latine:Age.L &  0.1972 &   [-0.3350, 0.7183] &     0.7712 &        0.7093 & 0.3512 \\
28 &        b\_Multiracial:Age.L & -0.0398 &   [-1.2559, 1.1419] &     0.5302 &        0.4932 & 0.3286 \\
29 &              b\_Other:Age.L & -0.0304 &   [-1.1689, 1.1053] &     0.5245 &        0.4874 & 0.3203 \\
30 &          b\_EastAsian:Age.Q & -0.4255 &   [-1.3038, 0.3890] &     0.8489 &        0.8225 & 0.6182 \\
31 &              b\_Black:Age.Q & -0.4501 &   [-1.0630, 0.1464] &     0.9335 &        0.9077 & 0.6903 \\
32 &         b\_SouthAsian:Age.Q & -0.0561 &   [-0.6773, 0.5760] &     0.5667 &        0.5066 & 0.2185 \\
33 &         b\_Indigenous:Age.Q & -0.1658 &   [-1.7826, 1.4614] &     0.5869 &        0.5619 & 0.4293 \\
34 &             b\_Latine:Age.Q & -0.3064 &   [-0.9294, 0.3036] &     0.8324 &        0.7907 & 0.5080 \\
35 &        b\_Multiracial:Age.Q & -0.5961 &   [-1.8650, 0.5356] &     0.8512 &        0.8312 & 0.6950 \\
36 &              b\_Other:Age.Q & -0.9584 &   [-2.3248, 0.2798] &     0.9319 &        0.9199 & 0.8439 \\
37 &    b\_EastAsian:PhasePhase2 &  0.8678 &    [0.1119, 1.6615] &     0.9878 &        0.9832 & 0.9298 \\
38 &        b\_Black:PhasePhase2 &  0.7228 &    [0.4026, 1.0400] &     1.0000 &        1.0000 & 0.9952 \\
39 &   b\_SouthAsian:PhasePhase2 & -0.0006 &   [-3.2901, 3.2067] &     0.5003 &        0.4809 & 0.3926 \\
40 &   b\_Indigenous:PhasePhase2 &  1.0285 &   [-0.3968, 2.7742] &     0.9208 &        0.9097 & 0.8370 \\
41 &       b\_Latine:PhasePhase2 &  0.5724 &    [0.2981, 0.8507] &     1.0000 &        0.9999 & 0.9747 \\
42 &  b\_Multiracial:PhasePhase2 & -0.0020 &   [-3.3493, 3.3883] &     0.5008 &        0.4838 & 0.3969 \\
43 &        b\_Other:PhasePhase2 & -0.9384 &   [-2.2867, 0.2083] &     0.9435 &        0.9330 & 0.8575 \\
44 &    b\_EastAsian:PhasePhase3 &  0.7048 &   [-0.0250, 1.4814] &     0.9711 &        0.9600 & 0.8600 \\
45 &        b\_Black:PhasePhase3 &  0.5190 &    [0.2630, 0.7769] &     0.9999 &        0.9998 & 0.9535 \\
46 &   b\_SouthAsian:PhasePhase3 &  1.1974 &    [0.5526, 1.8541] &     0.9999 &        0.9998 & 0.9976 \\
47 &   b\_Indigenous:PhasePhase3 &  0.4888 &   [-0.8743, 2.0475] &     0.7586 &        0.7352 & 0.6048 \\
48 &       b\_Latine:PhasePhase3 &  0.3633 &    [0.0838, 0.6435] &     0.9962 &        0.9872 & 0.6707 \\
49 &  b\_Multiracial:PhasePhase3 & -0.1308 &   [-1.7450, 1.3259] &     0.5720 &        0.5436 & 0.4066 \\
50 &        b\_Other:PhasePhase3 &  0.0001 &   [-3.2061, 3.2832] &     0.5001 &        0.4843 & 0.3931 \\
51 &    b\_EastAsian:Highschool- &  0.1652 &   [-0.7489, 1.0806] &     0.6473 &        0.6022 & 0.3836 \\
52 &        b\_Black:Highschool- &  0.0466 &   [-0.7349, 0.7995] &     0.5494 &        0.4962 & 0.2541 \\
53 &   b\_SouthAsian:Highschool- & -0.0433 &   [-0.8836, 0.8138] &     0.5423 &        0.4948 & 0.2721 \\
54 &   b\_Indigenous:Highschool- &  0.9707 &   [-0.4372, 2.7363] &     0.9104 &        0.8979 & 0.8210 \\
55 &       b\_Latine:Highschool- &  0.5825 &   [-0.2132, 1.3951] &     0.9222 &        0.9023 & 0.7499 \\
56 &  b\_Multiracial:Highschool- & -0.0633 &   [-1.7105, 1.4716] &     0.5302 &        0.5068 & 0.3752 \\
57 &        b\_Other:Highschool- &  0.6167 &   [-0.9791, 2.6817] &     0.7762 &        0.7591 & 0.6492 \\
58 &          b\_EastAsian:Other & -0.0067 &   [-3.0268, 3.1336] &     0.5022 &        0.4817 & 0.3887 \\
59 &              b\_Black:Other &  0.3003 &   [-0.8705, 1.5579] &     0.6955 &        0.6643 & 0.5005 \\
60 &         b\_SouthAsian:Other & -0.0866 &   [-1.4440, 1.2131] &     0.5513 &        0.5217 & 0.3682 \\
61 &         b\_Indigenous:Other &  0.2446 &   [-1.2971, 1.9356] &     0.6212 &        0.5976 & 0.4696 \\
62 &             b\_Latine:Other & -0.2369 &   [-1.7541, 1.1442] &     0.6337 &        0.6049 & 0.4670 \\
63 &        b\_Multiracial:Other &  0.0002 &   [-3.3189, 3.1498] &     0.5002 &        0.4826 & 0.3922 \\
64 &              b\_Other:Other &  0.0120 &   [-3.0902, 3.3868] &     0.5042 &        0.4857 & 0.3917 \\
\bottomrule
\end{tabular}

\caption{Summary of results for \emph{QS intersectional GE} model of expert-annotated data. These results show the estimated odds according to the model for the class in question versus the variable's reference class that the conversation being rated \emph{safe}. Note that, for gender \emph{Man} is the reference class, for race/ethnicity it is \emph{White}, for dataset phase it is \emph{Phase 1}, and for education it is \emph{College degree or above}. The suffix \textit{.L} represents the estimated linear effect of a variable (e.g., from Gen Z to Millenial); \textit{.Q} represents the estimated quadratic effect of a variable (e.g., from Gen Z to Gen X+); \textit{.C} represents the estimated cubic effect of a variable (e.g., from conversation severity of \textit{Benign} to \textit{Extreme}. \textit{Direction}, \textit{Significance}, and \textit{Large} correspond to probabilities of an effect's direction, practical significance, and magnitude being large, respectively.}
\label{tab:expert}
\label{tab:ad_sexit}
\end{table*}
\end{document}